\documentclass[submission, PhysCodeb,book]{SciPost}

\usepackage{tikz}
\usetikzlibrary{quantikz}
\usepackage{tikz-cd}
\tikzcdset{scale cd/.style={every label/.append style={scale=#1},
    cells={nodes={scale=#1}}}}
\usepackage{physics}
\usepackage[outline]{contour} 
\usetikzlibrary{intersections}
\usetikzlibrary{decorations.markings}
\usetikzlibrary{angles,quotes} 
\usetikzlibrary{calc}
\usetikzlibrary{3d}

\tikzset{>=latex} 
\colorlet{myred}{red!85!black}
\colorlet{myblue}{blue!80!black}
\colorlet{mycyan}{cyan!80!black}
\colorlet{mygreen}{green!70!black}
\colorlet{myorange}{orange!90!black!80}
\colorlet{mypurple}{red!50!blue!90!black!80}
\colorlet{mydarkred}{myred!80!black}
\colorlet{mydarkblue}{myblue!80!black}
\tikzstyle{xline}=[myblue,thick]

\tikzstyle{myarr}=[myblue!50,-{Latex[length=3,width=2]}]

\usepackage{lipsum}
\usepackage{adjustbox}
\usepackage{longtable}
\usepackage{colortbl}
\usetikzlibrary{arrows}
\usepackage{ctable}
\usepackage{makeidx}\makeindex
\tikzset{
operator/.append style={fill=purple!20},
my label/.append style={above right,xshift=0.3cm},
phase label/.append style={label position=above}
}
\makeatletter
 \DeclareRobustCommand\rvdots{%
\vbox{%
\baselineskip4\p@\lineskiplimit\z@%
 \kern-\p@%
 \hbox{.}\hbox{.}\hbox{.}%
}%
 }
\usepackage{subcaption}
\usepackage{graphics}
\graphicspath{ {images/} }
\usepackage{float}

\newcommand*{\gateStyle}[1]{{\textsf{\itshape #1}}}
\newcommand*{\hGate}{\gateStyle{H}}

\newrobustcmd{\B}{\bfseries}

\newcommand*{\circuitH}{\gate[style={fill=teal!20},label style=black]{\textnormal{\hGate{}}}}

\usepackage{yquant}

\usepackage{makecell}
\usepackage{tabularx}

\usepackage[utf8]{inputenc}
\usepackage{amssymb}
\usepackage{amsmath}
\usepackage{colortbl}
\usepackage{siunitx}
\usepackage{upgreek}
\usepackage{multirow}
\usepackage{hyperref}
\hypersetup{
    colorlinks=true,
    linkcolor=blue!70,
    citecolor=blue,
    filecolor=magenta,      
    urlcolor=blue,
}
\usepackage{algorithm}
\usepackage{algpseudocode}

\usepackage{tikz}
\usepackage{pgfplots}
\usepackage{csquotes}

\usepackage[english]{cleveref}

\pgfplotsset{compat=newest}
\usetikzlibrary{arrows}
\usepackage{pgfplots}

\definecolor{fluxcolor}{RGB}{204, 217, 255}
\definecolor{uwavecolor}{RGB}{244, 220, 222}
\definecolor{FandUwavecolor}{RGB}{231,244,224}
\definecolor{cavitycolor}{RGB}{232, 200, 244}

\crefname{lstlisting}{listing}{listings}
\Crefname{lstlisting}{Listing}{Listings}

\usepackage{xintexpr}
\usepackage{xintbinhex}

\usepackage{polexpr}[2018/02/16]

\makeatletter
\newcommand\FracBinToDecimal[1]{\romannumeral-`0%
  \expandafter\FracBin@ToDecimal\romannumeral0\xintraw{#1}%
}%
\def\FracBin@ToDecimal #1/#2[#3]{
  \ifnum#3<\z@
    \expandafter\@firstoftwo
  \else
    \expandafter\@secondoftwo
  \fi
  {\PolDecToString
     {\xintREZ{\xintiiMul{\xintBinToDec{#1}}{\xintiiPow{5}{-#3}}[#3]}}%
  }%
  {\xintiiMul{\xintBinToDec{#1}}{\xintiiPow{2}{#3}}}%
}%
\makeatother

\newcommand\test[1]{\[#1_b = \FracBinToDecimal{#1}_{10}\]}

\ExplSyntaxOn
\tl_new:N \l__bee_bindigits_tl
\seq_new:N \l__bee_quotients_seq
\seq_new:N \l__bee_remainders_seq

\NewDocumentCommand{\decbin}{ O{8} m }
 {
  \bee_decbin:nn { #1 } { #2 }
 }
\NewDocumentCommand{\bitcalc}{ s O{2} m }
 {
  \bee_bitcalc:nnn { #2 } { #3 } { \IfBooleanT{#1}{&} }
 }

\cs_new:Npn \bee_decbin:nn #1 #2
 {
  \int_compare:nTF { #2 >= 1 \prg_replicate:nn { #1 } { *2 } }
   {
    BAD! 
   }
   {
    \bee_print_decbin:nn { #1 } { #2 }
   }
 }
\cs_new_protected:Npn \bee_print_decbin:nn #1 #2
 {
  \tl_set:Nx \l__bee_bindigits_tl { \int_to_bin:n { #2 } }
  \prg_replicate:nn { #1 - \tl_count:N \l__bee_bindigits_tl }
   { \tl_put_left:Nn \l__bee_bindigits_tl { 0 } }
  $#2\sb{10}$
  \tl_map_inline:Nn \l__bee_bindigits_tl { \,\fbox{##1} }
 }

\cs_new_protected:Npn \bee_bitcalc:nnn #1 #2 #3
 {
  #2\sb{10}#3\to
  \seq_clear:N \l__bee_quotients_seq
  \seq_clear:N \l__bee_remainders_seq
  \bee_bitcalc_aux:nn { #1 } { #2 }
  \mathrm{ \str_upper_case:f { \int_to_base:nn { #2 } { #1 } }\sb{ #1 } }
 }
\cs_new_protected:Npn \bee_bitcalc_aux:nn #1 #2
 {
  \int_compare:nTF { #2 > 0 }
   {
    \seq_put_right:Nn \l__bee_quotients_seq { #1)\!\underline{\,\, #2 } }
    \seq_put_right:Nx \l__bee_remainders_seq { \int_mod:nn { #2 } { #1 } }
    \bee_bitcalc_aux:nx { #1 } { \int_div_truncate:nn { #2 } { #1 } }
   }
   {
    \bee_print_computation:
   }
 }
\cs_generate_variant:Nn \bee_bitcalc_aux:nn { nx }
\cs_new_protected:Npn \bee_print_computation:
 {
  \left\{
   \begin{array}{r}\seq_use:Nn \l__bee_quotients_seq { \\ } \end{array}
   \middle\uparrow
   \begin{array}{c}\seq_use:Nn \l__bee_remainders_seq { \\ } \end{array}
   \right\}
 }
\ExplSyntaxOff

\usepackage{xintfrac, xinttools}
\usepackage{polexpr}[2018/02/16]

\newcommand\MiniConvert[1]{\ifcase #1
  0\or 1\or 2\or 3\or 4\or 5\or 6\or 7\or 8\or 9\or A\or B\or C\or D\or E\or
  F\or G\or H\or I\or J\or K\or L\or M\or N\or O\or P\or Q\or R\or S\or T\or
  U\or V\or W\or X\or Y\or Z\else\ERROR\fi}%
\newcommand\ConvertitEnBaseB[3][25]{
     \def\ConvertiDots{\dots}%
  \noindent In base #3: \np{#2}.\par
  \def\Converti{0,}
  \edef\ConvertitNombre{\xintRaw{#2}}%
  \xintiloop[1+1]
  \edef\ConvertitBFoisNombre{\xintMul{#3}{\ConvertitNombre}}%
  \edef\ConvertitBFoisNombrePartieInt
      {\xintTTrunc{\ConvertitBFoisNombre}}%
  \edef\ConvertitBFoisNombrePartieFrac
      {\xintTFrac{\ConvertitBFoisNombre}}%
  $#3\times\np{\PolDecToString{\ConvertitNombre}}
             = \boxed{\ConvertitBFoisNombrePartieInt} +
               \np{\PolDecToString{\ConvertitBFoisNombrePartieFrac}}$
  \hfill
  \llap{${}\longrightarrow{}$\MiniConvert\ConvertitBFoisNombrePartieInt}\par
  \edef\Converti{\Converti\MiniConvert{\ConvertitBFoisNombrePartieInt}}%
  \let\ConvertitNombre\ConvertitBFoisNombrePartieFrac
  \xintifZero{\ConvertitNombre}
    {\xintbreakiloopanddo\let\ConvertiDots\empty.}%
    {}%
  \ifnum#1>\xintiloopindex\space
  \repeat
  \noindent\mbox{}\hfill$\np{#2}=[$\Converti\ConvertiDots$]_{#3}$\par
}
\newcommand\ConvertitFracEnBaseB[3][25]{%
     \def\ConvertiDots{\dots}%
  \edef\ConvertitNombre{\xintIrr{#2}}%
  \def\Converti{0,}
  \noindent In base #3: \ConvertitNombre.\par
  \xintiloop[1+1]
  \edef\ConvertitBFoisNombre{\xintMul{#3}{\ConvertitNombre}}%
  \edef\ConvertitBFoisNombrePartieInt
      {\xintTTrunc{\ConvertitBFoisNombre}}%
  \edef\ConvertitBFoisNombrePartieFrac
      {\xintTFrac{\ConvertitBFoisNombre}}
  $#3\times\xintFrac{\xintRawWithZeros\ConvertitNombre}
             = \boxed{\ConvertitBFoisNombrePartieInt} +
               \xintFrac{\xintRawWithZeros\ConvertitBFoisNombrePartieFrac}$\par
  \hfill
  \llap{${}\longrightarrow{}$\MiniConvert\ConvertitBFoisNombrePartieInt}\par
  \edef\Converti{\Converti\MiniConvert{\ConvertitBFoisNombrePartieInt}}%
  \let\ConvertitNombre\ConvertitBFoisNombrePartieFrac
  \xintifZero{\ConvertitNombre}
    {\xintbreakiloopanddo\let\ConvertiDots\empty.}%
    {}%
  \ifnum#1>\xintiloopindex\space
  \repeat
  \noindent\mbox{}\hfill$\xintFrac{#2}=[$\Converti\ConvertiDots$]_{#3}$\par}%

\usepackage[customcolors]{hf-tikz}

\tikzset{style green/.style={
    set fill color=green!50!lime!60,
    set border color=white,
  },
  style cyan/.style={
    set fill color=cyan!90!blue!60,
    set border color=white,
  },
  style orange/.style={
    set fill color=orange!80!red!60,
    set border color=white,
  },
  hor/.style={
    above left offset={-0.15,0.31},
    below right offset={0.15,-0.125},
    #1
  },
  ver/.style={
    above left offset={-0.1,0.3},
    below right offset={0.15,-0.15},
    #1
  }
}

\usepackage{listings} 
\definecolor{commentsColor}{rgb}{0.497495, 0.497587, 0.497464}
\definecolor{keywordsColor}{rgb}{0.000000, 0.000000, 0.635294}
\definecolor{stringColor}{rgb}{0.558215, 0.000000, 0.135316}

\usepackage{musixtex}
\usepackage{wasysym}
\usepackage{harmony}

\usepackage[font=bf,labelsep=space]{caption}

\AtBeginDocument{
	\IfFileExists{lmodern.sty}{
		\RequirePackage{lmodern}
	}{
          \IfFileExists{type1ec.sty}{}{
            \iftoggle{@noarxiv}{}{%
              \ClassError{quantumarticle}{Either the lmodern or the cm-super package are required by quantumarticle in order to produce high quality pdf output. Please install the lmodern package, or, if this is not possible, the cm-super package. You can turn off this error message by giving the noarxiv class option}{}
            }
          }
	}
	\IfFileExists{bbm.sty}{
		\RequirePackage{bbm}
		
	}{
		\IfFileExists{dsfont.sty}{
			\RequirePackage{dsfont}
			
		}{
			
		}
	}
	\RequirePackage{xcolor}
	\definecolor{quantumviolet}{HTML}{53257F} 
	\definecolor{quantumgray}{HTML}{555555} 
}

\begin{document}
\begin{center}{\Large \textbf{
A quantum Fourier transform (QFT) based note detection algorithm.\\
}}\end{center}

\markboth{Alqasemi et al.}{A quantum Fourier transform (QFT) based note detection algorithm.}
\title{A quantum Fourier transform (QFT) based note detection algorithm.}
\begin{center}
\eighthnote ~~~ \halfnote ~~~ \twonotes ~~~\\
\raisebox{-0.2em}{\textcolor{fluxcolor}{\rule{1em}{1em}}} Maryam Alqasemi \textsuperscript{1$\star$} \url{malqase1@jh.edu} ${}^\star$ {\small \sf JHU.}\\ 
\raisebox{-0.2em}{\textcolor{uwavecolor}{\rule{1em}{1em}}} Jacob Hammond \textsuperscript{2$\divideontimes$} \url{jhammo32@jh.edu} ${}^\divideontimes$ {\small \sf JHU.}\\ 
\raisebox{-0.2em}{\textcolor{FandUwavecolor}{\rule{1em}{1em}}} Shlomo Kashani \textsuperscript{3$\circledast$} \url{skashan2@jh.edu} ${}^\circledast$ {\small \sf JHU} \href{https://scholar.google.com.mx/citations?user=bM0LGgcAAAAJ&hl}{Google scholar.}\\   
\end{center}

\section*{Abstract}
\index{QFT}
In quantum information processing (QIP), the quantum Fourier transform (QFT) has a plethora of applications \cite{camps2020quantum} \cite{vorobyov2020quantum} \cite{mastriani2020fouriers}: Shor's algorithm and phase estimation are just a few well-known examples.
Shor's quantum factorization algorithm, one of the most widely quoted quantum algorithms \cite{nielsen00} \cite{rlz2016} \cite{skosana2021demonstration} relies heavily on the QFT and efficiently finds integer prime factors of large numbers on quantum computers \cite{nielsen00}. This seminal ground-breaking design for quantum algorithms has triggered a cascade of viable alternatives to previously unsolvable problems on a classical computer that are potentially superior and can run in polynomial time.
In this work we examine the QFT's structure and implementation for the creation of a quantum music note detection algorithm both on a simulated and a real quantum computer. Though formal approaches \cite{kopczyk2018quantum}\cite{camps2020quantum}\cite{hietala2019verified}\cite{r2018qwire} exist for the verification of quantum algorithms, in this study we limit ourselves to a simpler, symbolic representation which we validate using the symbolic SymPy \cite{qcm}\cite{Muoz2016QUANTUMAW} package which symbolically replicates quantum computing processes. The algorithm is then implemented as a quantum circuit, using IBM's qiskit \cite{portugal2022basic} library and finally period detection is exemplified on an actual single musical tone using a varying number of qubits.
\index{SymPy}
\index{DFT}
\index{Shor's algorithm}

\raisebox{-0.2em}{\textcolor{cavitycolor}{\rule{1em}{1em}}} \textbf{Keywords:} \AAcht, The quantum Fourier transform, QFT, DFT, quantum algorithms, quantum simulation, qiskit, SymPy, Cog, formal quantum algorithm verification. 
\maketitle
\definecolor{arsenic}{rgb}{0.23, 0.27, 0.29}
\vspace{2pt}
\noindent\rule{\textwidth}{1pt}
{
  \hypersetup{linkcolor=arsenic}
  \tableofcontents
}
\noindent\rule{\textwidth}{1pt}
\vspace{2pt}


\section{INTRODUCTION} \label{sec:introduction}
\index{FFT}

Classical computing progressed as a means of rapidly tackling real-world problems. The speed and capabilities of these computers have continually increased over the past few decades, enabling them to be used in virtually every aspect of our lives. However, as transistors shrink to the size of individual atoms \cite{atoms2017}, Moore's Law's projected exponential growth may stall. 

Quantum computing is one proposed answer to this challenge. Benioff \cite{benioff1997models} was the first to propose the use of quantum systems for information processing. Deutsch described quantum computers in 1985, which leverage the superposition of many particle states to achieve massive parallelism \cite{nielsen00}. Researchers have also investigated the prospect of tackling specific classes of problems more effectively than conventional computers can \cite{nielsen00}. These theoretical possibilities have piqued the curiosity of many researchers in the realisation of quantum hardware \cite{nielsen00} \cite{vorobyov2020quantum} \cite{mastriani2020fouriers} \cite{kopczyk2018quantum}\cite{camps2020quantum}.

The quantum world is fundamentally different from the classical one. A particular, significant implication is that, for certain classes of problems, computations can be performed more efficiently if the information is processed using quantum rather than conventional methods. In mathematics, convolution is an operation acting on two functions that produces a third function expressing how the shape of one is modified by the other. In classical computing, convolution is known to be explicitly difficult with a computational complexity that is exponential to the number of input functions \cite{nielsen00}. In linear, time-invariant systems where the input and output relationship is described by a convolution, this complexity can be reduced by using the Fourier transform. The Fourier transform of the convolution of two functions is simply the product of the Fourier Transforms of the functions. 

On the importance of the QFT one can learn from the work of Mastriani's \textit{Fourier's quantum information processing} \cite{mastriani2020fouriers}:

\begin{displayquote}
\raisebox{-0.2em}{\textcolor{cavitycolor}{\rule{1em}{1em}}}\textit{"We demonstrate that quantum information processing (QIP) completely rests on quantum Fourier transform (QFT), and 190 years after his death, the work of Jean-Baptiste Joseph Fourier is more present than ever in Physics, constituting the heart of QIP, and showing the spectral nature of quantum entanglement, quantum teleportation, and quantum secret sharing." \cite{mastriani2020fouriers}} \raisebox{-0.2em}{\textcolor{cavitycolor}{\rule{1em}{1em}}}
\end{displayquote}

The Fast Fourier Transform (FFT) is a frequently used technique for calculating the discrete Fourier transforms (DFT) of complex or real-valued data sequences \cite{vorobyov2020quantum}. The transformed data can be partitioned into the numerous pure frequencies that comprise it (\ref{fg0023}), a technique that is applicable to a wide variety of applications, including computational physics as well as machine and deep learning. Using the FFT in this case reduces computational complexity from $O\left(N^{2}\right)$ to $O(N \log N)$ \cite{mastriani2020fouriers}. This speedup is powerful and crucial to real-time processing of data. 


\subsection{Motivation} \label{sec:introduction}
Due to the fact that the QFT is fundamental to a large number of quantum algorithms \cite{vorobyov2020quantum}, the motivation of this paper is to offer a proof-of-concept by utilizing the QFT for discrete signal processing and ultimately showing how a quantum computer can provide reduced complexity over classical computers. 

\begin{figure}[h]
\begin{tikzpicture}
\begin{axis}[
    set layers=standard,
    domain=0:20,
    samples y=1,
    view={40}{20},
    hide axis,
    unit vector ratio*=1 2 1,
    xtick=\empty, ytick=\empty, ztick=\empty,
    clip=false
]
\def\sumcurve{0}
\pgfplotsinvokeforeach{0.5,1.5,...,5.5}{
    \draw [on layer=background, gray!20] (axis cs:0,#1,0) -- (axis cs:20,#1,0);
    \addplot3 [on layer=main, blue!30, smooth, samples=300]
      (x,#1,{sin(#1*x*(157))/(#1*2)});
    \addplot3 [on layer=axis foreground, very thick, blue,ycomb, samples=2]
      (10.5,#1,{1/(#1*2)});
    \xdef\sumcurve{\sumcurve + sin(#1*x*(157))/(#1*2)}
}
\addplot3 [red, samples=300] (x,0,{\sumcurve});
\draw [on layer=axis foreground]  (axis cs:0,0,0) -- (axis cs:20,0,0);
\draw (axis cs:20.5,0.25,0) -- (axis cs:20.5,5.5,0);
\end{axis}
\end{tikzpicture}
\caption{Time and frequency domain representations of a square function using the Fourier transform.}
\label{fg0023}
\end{figure}
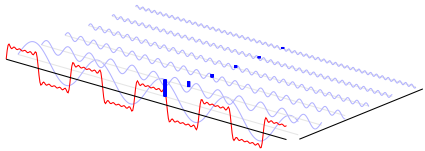

To show this in a tangible way, we will be using sampled audio files and use the QFT to determine the frequencies (\ref{fg0023}) of the audio. Since musical notes are assigned specific frequencies, a sampled audio recording can be represented as a convolution of one or many frequencies or notes. 
We can demonstrate that the quantum circuit can be encoded with the digital signal samples of some audio recording and efficiently determine which notes or frequencies are being played. We hope this contribution can be extended in future research to provide a template for reduced-complexity signal processing by using quantum computers.

%


\section{THE QFT} \label{sec:qft}

In this section, we present the theory of the QFT, and then we focus on how to use a formal approach for proofs of theorems to verify the correctness of a quantum algorithm in general, and the QFT in particular \cite{kopczyk2018quantum}\cite{camps2020quantum}\cite{hietala2019verified}\cite{r2018qwire}.

\subsection{The DFT} \label{sec:dft}

This section delves into the DFT and its properties in further depth. The Fourier transform turns a time-domain function into the frequencies that it is made up of in the frequency domain, transforming a list of evenly spaced function samples into a list of coefficients for a finite sequence of complex sinusoids, ordered by frequency.

The DFT assumes a discrete signal $\boldsymbol{f}=\left[f_{0}, \ldots, f_{N-1}\right]$ and computes it's frequency content $\boldsymbol{c}=\left[c_{0}, \ldots, c_{N-1}\right]$ by evaluating:

\begin{equation}
c_{k}=\sum_{j=0}^{N-1} f_{j} \omega^{j k}
\end{equation}

Where:
\begin{equation}
\label{eq002}
\omega=e^{-i \frac{2 \pi}{N}}
\end{equation}

Thus it maps a vector whose representation is $f=\left(f_{0}, f_{1}, \ldots, f_{N-1}\right)^{T}$ to a new vector $c=\left(c_{0}, c_{1}, \ldots, c_{N-1}\right)^{T}$ as follows:

\begin{equation}
c_{k}=\frac{1}{\sqrt{N}} \sum_{j=0}^{N-1} f_{j} e^{-2 \pi i \frac{k j}{N}}
\end{equation}

The inverse DFT is defined as:
\begin{equation}
f_{j}=\frac{1}{N} \sum_{k=0}^{N-1} C_{k} \bar{\omega}^{k j}
\end{equation}

\subsubsection{nth roots of unity}

Given $n \in \mathbb{N}$, the $n$-th roots of a complex number $\alpha$ are the solutions to the equation $z^{n}=\alpha$. When $\alpha=1$ then notion of roots of unity comes to fore and hence for any positive integer j:

\begin{equation}
\label{eq0012}
z=e^{i \frac{2 k \pi}{j}}
\end{equation}

Therefore \eqref{eq0012} is an $n$th root of 1 for any integer $k$, which we call an \textit{$n$th root of unity}. By letting:
\begin{equation}
\omega_{n}=e^{-i \frac{2 \pi}{n}}
\end{equation}

Then for any integer \textit{$k$}:
\begin{equation}
\omega_{n}^{k}=e^{-i \frac{2 k \pi}{n}}
\end{equation}

\begin{figure}[h]
        \centering
        \begin{tikzpicture}
        \pgfmathsetmacro{\n}{4}
        \begin{axis}[axis equal,
        axis lines=center,
        xlabel=$\Re$,
        xtick={-1.5,-0.5,0,0.5,1.5},
        ytick={0,0.5},
        xmax=1.5,
        xmin=-1.5,
        ymax=1.5,
        ymin=-1.5,
        ylabel=$\Im$,
        samples=10,
        disabledatascaling]
        \draw[help lines, black] (0,0) circle (1);
        \foreach \t in {1,...,\n} {
        \edef\temp{\noexpand
        \node[fill=blue, circle, draw=blue, scale=0.5] at ( {cos((360*\t)/\n)}, {sin((360*\t)/\n)} ) {};
         }\temp}
         \foreach \t in {1,...,\n} {
         \edef\temp{\noexpand
         \node[fill=white, circle, draw=none, scale=0.7] at ( {1.25*cos((360*\t)/\n)}, {1.25*sin((360*\t)/\n)} ) {$w_{\t}$};             
          }\temp}
        \end{axis}
        \end{tikzpicture}
        \caption{$n^{th}$ roots of unity on $\mathbb{C}$-plane}
        \label{fg001}
\end{figure}
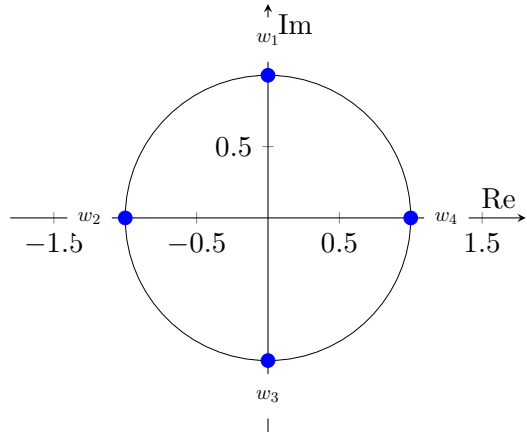

The distinct roots of unity are (see also (\ref{fg001})):
\begin{equation}
1, \omega_{n}, \omega_{n}^{2}, \ldots, \omega_{n}^{n-1}
\end{equation}

\subsubsection{The matrix form of the DFT}

The matrix $F_{N}$ form of the DFT is commonly used to calculate the DFT of a signal ${x}=\left(x_{0}, x_{1},\ldots x_{N-1}\right) \in \mathbb{C}^{N}$, and has as it's elements the numbers $w^{k \cdot n}$, where $k$ and $n, 0 \leq k, n \leq N-1$, are the row and column indices and $w=e^{-2 \pi i / N}$. For instance, as we saw earlier for the non-matrix DFT form, for N=4:

\begin{equation}
\begin{array}{r}
\omega_{4}=e^{-i \frac{2 \pi}{4}}=e^{-i \frac{\pi}{2}}=-i\\
\mathrm{~W}_{4}^{0}=1, W_{4}^{1}=-i, W_{4}^{2}=-1, W_{4}^{3}=i
\end{array}
\end{equation}

So that:
\begin{equation}
\label{eq022}
\boldsymbol{{F}_{4}}=\left[\begin{array}{cccc}
1 & 1 & 1 & 1 \\
1 & \omega^{1} & \omega^{2} & \omega^{3} \\
1 & \omega^{2} & \omega^{4} & \omega^{6} \\
1 & \omega^{3} & \omega^{6} & \omega^{9}
\end{array}\right]=\left[\begin{array}{cccc}
1 & 1 & 1 & 1 \\
1 & -i & -1 & i \\
1 & -1 & 1 & -1 \\
1 & i & -1 & -i
\end{array}\right]
\end{equation}

Thus using the same reasoning, the first four DFT matrices can be generated and are given by:
\begin{equation}
\label{eq023}
\boldsymbol{F_{1}}=[1], \quad F_{2}=\frac{1}{\sqrt{2}}\left[\begin{array}{rr}
1 & 1 \\
1 & -1
\end{array}\right], \quad \boldsymbol{F_{3}}=\frac{1}{\sqrt{3}}\left[\begin{array}{ccc}
1 & 1 & 1 \\
1 & \omega_{3} & \omega_{3}^{2} \\
1 & \omega_{3}^{2} & \omega_{3}
\end{array}\right], \quad \boldsymbol{F_{4}}=\frac{1}{2}\left[\begin{array}{rrrr}
1 & 1 & 1 & 1 \\
1 & -\imath & -1 & \imath \\
1 & -1 & 1 & -1 \\
1 & \imath & -1 & -\imath
\end{array}\right]
\end{equation}

\subsubsection{Numerical examples: DFT}
To gain a better understanding of the inner workings of the DFT, we present several examples of manually computing the DFT in both it's regular and matrix forms.

\begin{enumerate}
\item N=2 and using \eqref{eq023} the DFT for the signal $\boldsymbol{x}=(1,2)$.\\
When $N=2$ we have $w=e^{-2 \pi i / 2}=e^{-\pi i}=-1$ and $\boldsymbol{F}_{2}=H=\left[\begin{array}{rr}1 & 1 \\ 1 & -1\end{array}\right]$ so the DFT is:

\begin{equation}
\boldsymbol{X}=\boldsymbol{F}_{2} \boldsymbol{x}=\left[\begin{array}{rr}
1 & 1 \\
1 & -1
\end{array}\right] \cdot\left[\begin{array}{l}
1 \\
2
\end{array}\right]=\left[\begin{array}{r}
3 \\
-1
\end{array}\right]
\end{equation}

\item N=4, the DFT for the signal $\boldsymbol{x}=(1,2,0,0)$.\\
When $N=4$ and using \eqref{eq022} we can obtain the DFT by matrix multiplication: 

\begin{equation}
\boldsymbol{X}=\boldsymbol{F}_{4} \boldsymbol{x}=\left[\begin{array}{rrrr}
1 & 1 & 1 & 1 \\
1 & -i & -1 & i \\
1 & -1 & 1 & -1 \\
1 & i & -1 & -i
\end{array}\right] \cdot\left[\begin{array}{l}
1 \\
2 \\
0 \\
0
\end{array}\right]=\left[\begin{array}{c}
3 \\
1-2 i \\
-1 \\
1+2 i
\end{array}\right]
\end{equation}

\end{enumerate}

\subsection{FROM THE DFT TO THE QFT}\label{dft2qft}

In quantum computing, the QFT \eqref{eq:qft} is the quantum analogue of the DFT. The QFT can be performed efficiently on a quantum computer using exponentially less gates than is required to compute classically. In QIP, the QFT is a Hadamard transform generalisation and both are quite similar, with the exception that the QFT introduces a phase \ref{eq:phase}.

\begin{enumerate}

\item[$\circ$] \textbf{The Quantum Fourier transformation}:
The QFT is a basis transformation described by the following unitary evolution. Let $N=2^{n}$ and $|k\rangle,|j\rangle$ denote computational basis vectors. Then:
\begin{equation}
|k\rangle \stackrel{\mathrm{QFT}}{\longmapsto} \frac{1}{\sqrt{2^{n}}} \sum_{k=0}^{2^{n}-1} \omega^{j k}|j\rangle
\end{equation}

Where $\omega$ is the $N$ th root of unity:
\begin{equation}
\omega:=e^{2 \pi i / 2^{n}}
\end{equation}

We are going to adopt the notation used in \cite{camps2020quantum}. Given a quantum state:
\begin{equation}
\label{eq:qft}
\psi\rangle=\sum_{j=0}^{N-1} a_{j}|j\rangle=\left(\begin{array}{c}
a_{0} \\
\vdots \\
a_{N-1}
\end{array}\right)
\end{equation}
The QFT can be computed as follows:
\begin{equation}
F|\psi\rangle=\sum_{k=0}^{N-1} b_{k}|k\rangle
\end{equation}

Where:
\begin{equation}
b_{k}=\frac{1}{\sqrt{N}} \sum_{j=0}^{N-1} a_{j} \mathrm{e}^{2 \pi \mathrm{i} j k / N}
\end{equation}

\end{enumerate}


\subsubsection{Relative phase shift}

Imagine two states $|\hat{\Theta}\rangle$ and $|\hat{\Phi}\rangle$ which differ only by a relative phase shift and thus, if measured, the phases will be insignificant and therefore the phase of the given
state \textbf{does not effect the probability of measuring that state}. For instance, after a measurement, these two states are indistinguishable: 
\begin{equation}
|\hat{\Phi}\rangle=\left(\begin{array}{c}\beta_{0} \\ -i \beta_{1} \\ -\beta_{2} \\ i \beta_{3}\end{array}\right)
\end{equation}
\begin{equation}
|\hat{\Theta}\rangle=\left(\begin{array}{c}\beta_{0} \\ \beta_{1} \\ \beta_{2} \\ \beta_{3}\end{array}\right)
\end{equation}

However, as we shall see later, if we apply the QFT on the states $|\hat{\Phi}\rangle$ and $|\hat{\Theta}\rangle$ measure their Fourier Transform, this action \textbf{will disclose the relative phases.}
\begin{equation}
Q F T_{4} \frac{1}{2}\left(\begin{array}{l}
1 \\
1 \\
1 \\
1
\end{array}\right)=\left(\begin{array}{l}
1 \\
0 \\
0 \\
0
\end{array}\right)
\end{equation}
\begin{equation}
Q_{F} T_{4} \frac{1}{2}\left(\begin{array}{c}
1 \\
i \\
-1 \\
-i
\end{array}\right)=\left(\begin{array}{l}
0 \\
1 \\
0 \\
0
\end{array}\right)
\end{equation}

\subsubsection{Binary number representation}

It is critical to maintain the ordering of qubits during creation, access to results, and conversion of binary strings. 
When qubits are introduced to a circuit, they are added left to right (as in a binary string), beginning with the highest-order qubit and ending with the lowest-order qubit.

Two tensor-ed states are denoted in Dirac notation by the symbol $|x, y\rangle$, for instance, $|0,1\rangle$.
Additionally, the most important bit appears \textbf{first} in this format.
This type of state is also stated in decimals \eqref{eq:dec0099} .

\begin{equation}
\label{eq:dec0099}
\underbrace{|0\rangle}_{\text {MSB}} \otimes \cdots \otimes \otimes \cdots \otimes \underbrace{|0\rangle}_{\text {LSB}}
\end{equation}

An example follows, wherein bit-strings are interpreted as binary integers:
\begin{equation}
|1\rangle \otimes|0\rangle \otimes|1\rangle=|101\rangle
\end{equation}

In the QFT, as we shall shortly show, it is common practice to use a binary representation of decimal numbers. For instance the binary representation for the number 17 is:
\[
\bitcalc{17}
\]
or for instance: 
\begin{equation}
11001.01_{b}=25.25_{10}
\end{equation}

\begin{equation}
0.001_{b}=0.125_{10}
\end{equation}

\test{10.1101}

In our case, in the Dirac notation, a 3-qubit state may be represented as follows: 
\begin{equation}
\left|\Psi\right\rangle=\frac{1}{\sqrt{8}}(|000\rangle+|001\rangle+\ldots+|111\rangle)=\frac{1}{\sqrt{8}}(|1\rangle+|2\rangle+\ldots|7\rangle)
\end{equation}

%
%

\section{FORMAL VERIFICATION OF THE QFT ALGORITHM}

In this section, we shall briefly examine approaches for the formal verification of quantum algorithms.
Quantum program accuracy is critical, and mathematicians have traditionally used \textit{theorem proving techniques} to formalize and verify conjectures and theorems \cite{hietala2019verified}.

Quantum information scientists will face a common dilemma \cite{boender2015formalization}\cite{r2018qwire}) as quantum computing develops from theory to practice: how can they be confident that their code performs what they wish it to do? For example, the authors propose \textit{"a novel framework for modelling and evaluating quantum protocols and their implementations using the proof assistant Coq"} in formalization of Quantum Protocols using Coq \cite{boender2015formalization}.This quantum programming difficulty presents an opportunity for formal techniques \cite{r2018qwire}. 

The Coq proof assistant is a mathematical programming language that is primarily used to ensure that algorithms are accurate \cite{boender2015formalization}.It's a general-purpose environment for creating mathematical proofs that's been enhanced by a number of authors \cite{kopczyk2018quantum}\cite{camps2020quantum}\cite{hietala2019verified}\cite{r2018qwire} to support quantum information processing proofs. 

Formal approaches can be used to verify that the code implementing a quantum algorithm accomplishes what it should for all potential inputs and configurations in advance. In this study, we limit ourselves to a simpler, symbolic approach
for representing and reasoning about the quantum QFT circuit, which we then validate using the Pythonic \textbf{SymPy} package \cite{sympy2012}. Symbolic computing  is used to manipulate symbols. and solve mathematical equations and formulas in order to get results that are mathematically correct. This method is based on a \textbf{symbolic depiction of quantum states and quantum gates using symbols}, rather than numbers.

\subsection{Symbolic representation of universal quantum gates}

A quantum operator can be any unitary matrix ($\left(U^{\dagger} U\right)=I$) of size $2^n$ and in the same way as traditional logic gates are referred to as operators, they are also referred to as gates. The Hadamard and the phase operator are not only unitary but also universal \cite{nielsen00}.
Here, we break down the Fourier transform into a set of universal gates made up of CNOT and one-qubit gates. This decomposition (\ref{decomp}) allows us to understand it's optimal realization on a real quantum computer. For instance, the Solovay-Kitaev is an algorithm that can approximate any single transformation by breaking it down into the basic gates that make it up \cite{sympy2012}.

\label{eq:phase}
\begin{displayquote}
\raisebox{-0.2em}{\textcolor{cavitycolor}{\rule{1em}{1em}}}\textit{"Any QC algorithm can be represented as a composition of Walsh-Hadamard transforms and associated conditional phase shifts" \cite{bowden2000universality}} \raisebox{-0.2em}{\textcolor{cavitycolor}{\rule{1em}{1em}}}
\end{displayquote}

From the \textbf{quantum optics} or hardware perspective, $H$ is realized by a half-silvered mirror (beam splitter) and $P(\omega)$ represents a phase shifter, as in a standard Mach-Zehnder interferometer \cite{cleve1997quantum} \cite{Klappenecker00computingwith} \cite{bowden2000universality}. In \cite{weinstein1999implementation} the authors implemented the QFT on a three bit nuclear magnetic resonance (NMR) quantum computer. Realizing the QFT using NMR works quite similarly \cite{bowden2000universality}. 

In the factorization equation \cite{Klappenecker00computingwith}  each factor is of the form $|0\rangle+e^{i \omega}|1\rangle$. This can be implemented using the application of two successive unitary operators. First, the Hadamard (\ref{fghadamard}), which is one of most commonly used quantum gates:

\begin{figure}[h]
\centering
\adjustbox{scale=1}{%
(i)
$H=\frac{1}{\sqrt{2}}\left[\begin{array}{rr}1 & 1 \\ 1 & -1\end{array}\right]$
(ii)
 \adjustbox{scale=1}{%
 \begin{tikzcd}
    \lstick{$\Psi$} & \circuitH & \qw \\
\end{tikzcd}
}
(iii)
}   $ (H) $ \\
\caption{The unitary gate for the 1-qubit Hadamard. The diagram is a commonly used \cite{nielsen00} schematic representation of the gate $H$ acting on a qubit in state $|\Psi\rangle$. (i) The Hadamard matrix (ii) The Hadamard gate (iii) The hadamard symbol. Applying the Hadamard on the quantum state $|0\rangle$ for instance, results in $H|0\rangle=\frac{1}{\sqrt{2}}(|0\rangle+|1\rangle)$.}
\end{figure}

\textit{The Hadamard has an intuitive interpretation around the Bloch sphere} \cite{nielsen00}: If the qubit is initialised in the ground state $|0\rangle$ and then transformed via a Hadamard operator, it's state is transformed from $|0\rangle$ to an equal superposition of 0 and 1 and thus rotates the quantum state vector from the north pole of Bloch's sphere to the equator. In the Fourier basis, this alters our computational basis states from  $|0\rangle$ and $|1\rangle$ to $|+\rangle$ and $|-\rangle$.

and the shift operator:
\begin{equation}
P(\omega)=\left[\begin{array}{cc}1 & 0 \\ 0 & e^{i \omega}\end{array}\right]
\end{equation}
together forming the following unitary:

\begin{equation}
P(\omega)[H|0\rangle]=\frac{1}{\sqrt{2}}\left(|0\rangle+e^{i \omega}|1\rangle\right)
\end{equation}
Therefore, the QFT can be realized using universal quantum gates \cite{nielsen00}. 

In the literature, there are several libraries for generating circuits from their respective unitary representation \cite{r2018qwire} and vice-versa. SymPy \cite{sympy2012} is one such library which encompasses algebra, calculus and a base module for \textit{symbolic quantum computing} in the \textit{Dirac} \cite{nielsen00} notation. 

\begin{displayquote}
\raisebox{-0.2em}{\textcolor{cavitycolor}{\rule{1em}{1em}}}\textit{"SymPy's quantum module now supports Hilbert spaces, states, spin systems, bras, kets, inner/outer products, operators, commutators, anticommutators, daggers, etc. in their most
abstract form." \cite{sympy2012}} \raisebox{-0.2em}{\textcolor{cavitycolor}{\rule{1em}{1em}}}
\end{displayquote}

Code listing (\ref{code0033}) depicts a typical usage, involving the $sympy.physics.quantum$ quantum package and the tensor product {$\Psi=\left(H_{0}|\phi\rangle\right) \otimes\left(H_{0}|\psi\rangle\right)$}. 
\begin{figure}[H]
\begin{lstlisting}[caption={},language=Python] 
from sympy import * 
from sympy.physics.quantum import TensorProduct

phi, psi = Ket('phi'), Ket('psi')
U_final=(TensorProduct(qapply(H(0)*phi),qapply(H(0)*psi)))
U_final
\end{lstlisting}
\caption{A typical usage involving the $sympy.physics.quantum$ quantum package. The tensor product $TensorProduct(psi1,psi2)$ results in \small{$\left(H_{0}|\phi\rangle\right) \otimes\left(H_{0}|\psi\rangle\right)$} \normalsize which is a symbolic rather than a numerical representation. \textit{Symbolically}, Sympy would represent the expression exactly the same way as a human would in the Dirac notation. From here stems the relative advantage of being able to use the same "language" both in proofs and in code.} 
\label{code0033}
\end{figure}

%


\subsection{Numerical examples: manually applying the QFT}\label{qft-man}
In this section, we present the results of manually applying the QFT's to a variety of quantum states. We repeat the exact same calculations later in sec (\ref{qft-sym}), except for using SymPy programmatically, to show that they are identical. 

\subsubsection{Numerical examples: \textit{1}-qubit QFT}


First, we note that the Hadamard matrix $H$ is \textbf{equivalent} to the 1-qubit QFT and it's inverse, that is, the QFT is reduced to applying a Hadamard gate:

\begin{equation}
F_{2^{1}}=H=\frac{1}{\sqrt{2}}\left[\begin{array}{rr}
1 & 1 \\
1 & -1
\end{array}\right]=H^{-1}=F_{2^{1}}^{-1}
\end{equation}
Also note that:
\begin{equation}
\begin{aligned}
\omega=e^{2 \pi i / N}=e^{2 \pi i / 2}=e^{i \pi}=-1 \\
\end{aligned}
\end{equation}
\begin{enumerate}
\item \raisebox{-0.2em}{\textcolor{cavitycolor}{\rule{1em}{1em}}} For $N=2^{1}=2$.
\begin{equation}
\label{eq787}
\begin{array}{rlc}
Q F T|x\rangle & =\quad \frac{1}{\sqrt{2}} \sum_{y=0}^{N-1} e^{2 \pi i x y / N}|y\rangle \\
& =\frac{1}{\sqrt{2}}\left[e^{2 \pi i x \cdot 0 / 2}|0\rangle+e^{2 \pi i x \cdot 1 / 2}|1\rangle\right] \\
& =\quad \frac{1}{\sqrt{2}}\left[|0\rangle+e^{\pi i x}|1\rangle\right]
\end{array}
\end{equation}

Now, we can apply \eqref{eq787} to the quantum states $|0\rangle$ and $|1\rangle$ respectively:
\begin{equation}
\begin{aligned}
&Q F T|0\rangle=\frac{1}{\sqrt{2}}\left[|0\rangle+e^{\pi i \cdot 0}|1\rangle\right]=\frac{1}{\sqrt{2}}(|0\rangle+|1\rangle)=|+\rangle \\
&Q F T|1\rangle=\frac{1}{\sqrt{2}}\left[|0\rangle+e^{\pi i \cdot 1}|1\rangle\right]=\frac{1}{\sqrt{2}}(|0\rangle-|1\rangle)=|-\rangle
\end{aligned}
\end{equation}

\item \raisebox{-0.2em}{\textcolor{cavitycolor}{\rule{1em}{1em}}} For the state $\psi=a_{0}|0\rangle+a_{1}|1\rangle$.\\
We’ve already seen that the QFT for N = 2 is the Hadamard transform $H=\frac{1}{\sqrt{2}}\left[\begin{array}{cc}1 & 1 \\ 1 & -1\end{array}\right]$. 
Applying the Hadamard transform to the quantum state $\psi$ we obtain the new state:
\begin{equation}
\frac{1}{\sqrt{2}}\left(a_{0}+a_{1}\right)|0\rangle+\frac{1}{\sqrt{2}}\left(a_{0}-a_{1}\right)|1\rangle=\tilde{a}_{0}|0\rangle+\tilde{a}_{1}|1\rangle
\label{fghadamard}
\end{equation}
\end{enumerate}

We now move on to the \textit{2}-qubit QFT. 
\subsubsection{Numerical examples: \textit{2}-qubit QFT.}

\begin{enumerate}

\item \raisebox{-0.2em}{\textcolor{cavitycolor}{\rule{1em}{1em}}} Given the following two quantum states in the Dirac notation.
\begin{equation}
|g\rangle=|0\rangle=\left(\begin{array}{l}
1 \\
0 \\
0 \\
0
\end{array}\right), \text { and }|h\rangle=|1\rangle=\left(\begin{array}{l}
0 \\
1 \\
0 \\
0
\end{array}\right)
\end{equation}

Then the corresponding Fourier transforms $Q F T_{4}$ applied to $|g\rangle$ and $Q F T_{4}$ applied to $|h\rangle$ are given by:

\begin{equation}
|\hat{g}\rangle=\frac{1}{2}\left(\begin{array}{cccc}
1 & 1 & 1 & 1 \\
1 & i & -1 & -i \\
1 & -1 & 1 & -1 \\
1 & -i & -1 & i
\end{array}\right)\left(\begin{array}{l}
1 \\
0 \\
0 \\
0
\end{array}\right)=\frac{1}{2}\left(\begin{array}{l}
1 \\
1 \\
1 \\
1
\end{array}\right)
\end{equation}

\begin{equation}
|\hat{h}\rangle=\frac{1}{2}\left(\begin{array}{cccc}
1 & 1 & 1 & 1 \\
1 & i & -1 & -i \\
1 & -1 & 1 & -1 \\
1 & -i & -1 & i
\end{array}\right)\left(\begin{array}{l}
0 \\
1 \\
0 \\
0
\end{array}\right)=\frac{1}{2}\left(\begin{array}{c}
1 \\
i \\
-1 \\
-i
\end{array}\right)
\end{equation}

\item\raisebox{-0.2em}{\textcolor{cavitycolor}{\rule{1em}{1em}}}  $QFT_{4}$ applied to the quantum state $|01\rangle$
\begin{equation}
\label{eq-angel}
F_{4}|01\rangle \equiv \frac{1}{2}\left[\begin{array}{cccc}
\tikzmarkin[hor=style cyan]{row} 1 & 1 & 1 & 1 \tikzmarkend{row}\\
1 & i & -1 & -i \\
1 & -1 & 1 & -1 \\
1 & -i & -1 & i
\end{array}\right]\left[\begin{array}{l}
0 \\
1 \\
0 \\
0
\end{array}\right]=\left[\begin{array}{c}
1 \\
i \\
-1 \\
-i
\end{array}\right] \equiv \frac{1}{2}(|00\rangle+i|01\rangle-|10\rangle-i|11\rangle)
\end{equation}

The operator \eqref{eq-angel} shows the QFT, acting on the input states by $0*\pi/2$ degrees in the \textcolor{cyan}{first row} and column, followed by $1*\pi/2$ degrees, $2*\pi/2$ degrees and $3*\pi/2$ degrees respectively, all of which are \textbf{multiples} of $\frac{\pi}{2}$.
\end{enumerate}

\subsection{Numerical examples: Symbolically applying the QFT}\label{qft-sym}
In this section, we present the results of programmatically applying the QFT's to a variety of quantum states. We repeat the exact same calculations as in sec (\ref{qft-man}) which were manually computed. Code listing (\ref{decomp}) illustrates the decomposition of the 2-qubit unitary \eqref{eq-angel} using SymPy. 

\begin{figure}[H]
\raisebox{-0.2em}{\textcolor{cavitycolor}{\rule{1em}{1em}}} Python:
\begin{lstlisting}[caption={},language=Python] 
from sympy import * 
state = Qubit('01')
fourier = QFT(0,2).decompose()
state, fourier
\end{lstlisting}
\raisebox{-0.2em}{\textcolor{cavitycolor}{\rule{1em}{1em}}} Symbolic decomposition: $SWAP_{0,1} H_{0} C_{0}\left(S_{1}\right) H_{1}$\\
\raisebox{-0.2em}{\textcolor{cavitycolor}{\rule{1em}{1em}}} Measurement: 
\begin{lstlisting}[caption={},language=Python] 
from sympy import * 
represent (qapply(fourier*state))
\end{lstlisting}
\begin{equation}
\label{sym-meas}
\left[\begin{array}{c}
\frac{1}{2} \\
\frac{i}{2} \\
-\frac{1}{2} \\
-\frac{i}{2}
\end{array}\right]
\end{equation}

\caption{A decomposition of the 2-qubit QFT unitary using SymPy. Note that the Symbolic measurement in \eqref{sym-meas} is identical to the one obtained manually in \eqref{eq-angel}.}  
\label{decomp}
\end{figure}

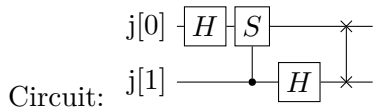
\begin{figure}[H]
\raisebox{-0.2em}{\textcolor{cavitycolor}{\rule{1em}{1em}}} Python:
\begin{lstlisting}[caption={},language=Python] 
from sympy import * 
circuit_plot(fourier, nqubits=2);
\end{lstlisting}
\label{circ-gen}
\raisebox{-0.2em}{\textcolor{cavitycolor}{\rule{1em}{1em}}} Circuit: 
\begin{tikzpicture}
\begin{yquant}
qubit j[2];
h j[0];
box {$S$} j[0] | j[1];
h j[1];
swap (j[0, 1]);
\end{yquant}
\end{tikzpicture}
\caption{A generation of the 2-qubit QFT circuit, from the respective unitary using SymPy.}  
\label{code0099}
\end{figure}

Furthermore, using Sympy we now generate the circuit out of the symbolic representation (\ref{code0099}). 

\section{DEMONSTRATING THE QFT USING QISKIT}\label{sec:qiskit_demo}
In this section, we utilize Qiskit to demonstrate the QFT. First, by building a single qubit system and scaling it up to n qubits and then use the QFT circuit to do conduct quantum measurements on a real quantum device. Finally, we demonstrate the relationship between the QFT and classical FFT function.
\newline
\newline
While the QFT can be implemented on any number of qubits, the complexity of the circuit implementation in Qiskit grows with more qubits. For the purpose of this study, we will present an example on three qubits which could be further extended up to n-qubits. 

\subsection{Qiskit Background}
Qiskit is an open-source SDK for working with quantum computers at the level of pulses, circuits, and application modules. Qiskit includes a comprehensive set of quantum gates and a variety of pre-built circuits so users at all levels can use Qiskit for research and application development. Specifically, Qiskit can be used to implement the QFT on a simulated quantum computer or on a cloud-based Quantum Computer.\cite{qiskit}

\subsection{Preparing the Quantum circuit}
To  build a QFT circuit, we begin with the simplest case, a 1-qubit system. As noted in section \ref{qft-man}, the 1-qubit QFT is equivalent to a Hadamard gate. In Qiskit, this is easily constructed using the code:

\begin{figure}[H]
\raisebox{-0.2em}{\textcolor{cavitycolor}{\rule{1em}{1em}}} Python:
\begin{lstlisting}[caption={},language=Python] 
from qiskit import QuantumCircuit
qc.h(0)
qc.draw()
\end{lstlisting}
\label{circ-gen}
\raisebox{-0.2em}{\textcolor{cavitycolor}{\rule{1em}{1em}}} Circuit: 
\begin{tikzpicture}
\begin{yquant}
qubit j[1];
h j[0];
\end{yquant}
\end{tikzpicture}
\caption{A generation of the 1-qubit QFT circuit using qiskit.}  
\label{qiskit1}
\end{figure}

Considering the basis $j=0,1$, we have:

\begin{equation}
\begin{aligned}
&Q F T|0\rangle=\frac{1}{\sqrt{2}}\left[|0\rangle+|1\rangle\right] \\
&Q F T|1\rangle=\frac{1}{\sqrt{2}}\left[|0\rangle-|1\rangle\right]
\end{aligned}
\end{equation}

We can use Qiskit to demonstrate the transformation on top of the Bloch sphere, where it transforms the computational basis on the two poles of the sphere to the Fourier basis on the \textbf{equatorial} of the sphere.

\begin{figure}[H]
\centering{
\includegraphics[scale=0.31]{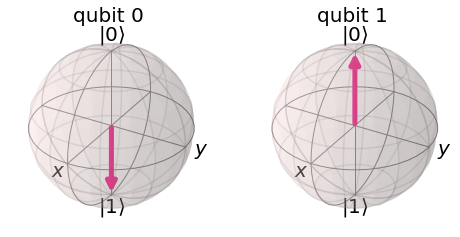}\\
\includegraphics[scale=0.31]{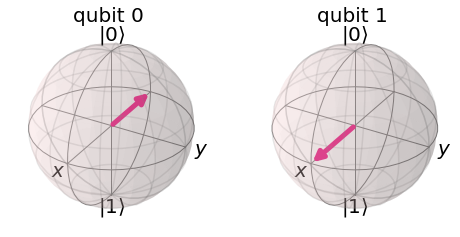}}
\caption{Bloch sphere representation of the QFT applied on the basis states.}
\end{figure}

We now consider a 2-qubit circuit (n=2): 
\begin{equation}
QFT |j\rangle =\frac{1}{2}(|0\rangle + e^{\pi i j} |1\rangle) \otimes (|0\rangle + e^{\pi i j/2} |1\rangle)
\end{equation}

A QFT applied to this system can be decomposed using the two quantum gates:
\begin{enumerate}
\item a Hadamard gate $\hat{H}|j\rangle =\frac{1}{\sqrt{2}} (|0\rangle + e^{\frac{2\pi i k_m}{2}}|1\rangle)$
\item a controlled phase gate $S_k|j\rangle= e^{2\pi i j/2^k}|j \rangle$
\end{enumerate}
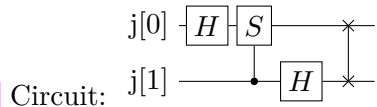
\begin{figure}[H]
\raisebox{-0.2em}{\textcolor{cavitycolor}{\rule{1em}{1em}}} Python:
\begin{lstlisting}[caption={},language=Python] 
from qiskit import QuantumCircuit
pi=np.pi
qc=QuantumCircuit(2)
qc.h(0)
qc.cp(pi/2,1,0)
qc.h(1)      
qc.draw()
\end{lstlisting}
\label{circ-gen}
\raisebox{-0.2em}{\textcolor{cavitycolor}{\rule{1em}{1em}}} Circuit: 
\begin{tikzpicture}
\begin{yquant}
qubit j[2];
h j[0];
box {$S$} j[0] | j[1];
h j[1];
swap (j[0, 1]);
\end{yquant}
\end{tikzpicture}
\caption{A generation of the 2-qubit QFT circuit using qiskit}  
\label{qiskit1}
\end{figure}

\subsection{Scaling of the Quantum Circuit as a function of N}
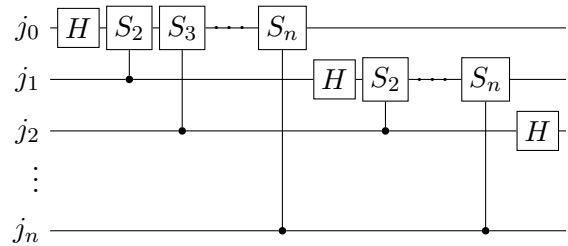
\begin{figure}[H]
\centering{
\yquantdefinebox{dots}[inner sep=0pt]{$\dots$}
\begin{tikzpicture}
\begin{yquant}
qubit {$j_0$} j0;
qubit {$j_1$} j1;
qubit {$j_2$} j2;
qubit {$\rvdots$} x3; discard x3;
qubit {$j_n$} j4;
h j0;
box {$S_2$} j0 | j1;
box {$S_3$} j0 | j2;
dots j0;
box {$S_n$} j0 | j4;
h j1;
box {$S_2$} j1 | j2;
dots j1;
box {$S_n$} j1 | j4;
h j2;
\end{yquant} 
\end{tikzpicture}
\caption{n-qubit QFT circuit}}
\label{nqcircuit}
\end{figure}

We have built two simple QFT circuits for n=1 and n=2. To Build a circuit that implements the QFT for an n-arbitrary number of qubits, we go back to the definition of the QFT in section \ref{dft2qft}.  For $N=2^n$, the QFT is defined as:
\begin{equation}
QFT |j\rangle =\frac{1}{\sqrt{2^{n}}} \sum_{k=0}^{2^{n}-1} e^{2 \pi i j k / 2^{n}}|k\rangle
\end{equation}
To better understand how the quantum circuit is built, we apply some simplifications. For $k$:
$$
|k\rangle =|k_1 k_2 ....k_n\rangle
$$
Where $k_n \in {0,1}$. Hence, we can write the binary expansion of $k$ in the following way:
\begin{equation}
k=2^{n-1}k_1 + 2^{n-2} k_2 + .....+ 2^{n-n}k_n=\sum_{m=1}^{n}2^{n-m}k_m
\end{equation}
Thus we have:
$$
QFT |j\rangle =\frac{1}{\sqrt{2^{n}}} \sum_{k=0}^{2^{n}-1} e^{2 \pi i j (\sum_{m=1}^{n}2^{n-m}k_m)/ 2^{n}}|k\rangle
$$
$$
QFT |j\rangle =\frac{1}{\sqrt{2^{n}}}\sum_{k_1=0}^{1}\sum_{k_2=0}^{1}\sum_{k_3=0}^{1}...\sum_{k_n=0}^{1}\Pi_{m=1}^{n} e^{\frac{2\pi i j k_m}{2^m}}|k_1 k_2 ....k_n\rangle
$$
Simplifying:
\begin{equation}
QFT |j\rangle =\frac{1}{\sqrt{2^{n}}}\otimes^{n}_{m=1} \sum_{k_m=0}^1 e^{\frac{2\pi i j k_m}{2^m}}|k_m\rangle
\end{equation}
\begin{equation} \label{eq:1}
QFT |j\rangle=\frac{1}{\sqrt{2^n}} \otimes_{m=1}^n(|0\rangle + e^{2 \pi i j/2^n} |1\rangle)
\end{equation}
The above formula can be used in building a quantum circuit that implements the QFT. Where we start by applying a Hadamard gate on the first qubit and then a series of a controlled phase gates. Similarly for the second qubit and third till the n-th qubit. \ref{nqcircuit} Shows an n-qubit circuit of this form.
\newline
However, building this circuit, gives up a reversed order of the derived form in \ref{eq:1}. Therefore we add up a swap gate at the end to get the correct QFT implementation.
 
\begin{figure}[H]
\raisebox{-0.2em}{\textcolor{cavitycolor}{\rule{1em}{1em}}} Python:
\begin{lstlisting}[caption={},language=Python] 
from qiskit import QuantumCircuit
import numpy as np
pi=np.pi

def QFTn(n):
    qcn=QuantumCircuit(n)
    for q in range(n):
        qcn.h(q)
        for qp1 in range(q+1,n):
            qcn.cp(pi/(2**(qp1-q)),qp1,q)       
    for q in range(n//2):
        qcn.swap(q, n-q-1)
    
    return qcn
\end{lstlisting}
\caption{A generation of a general QFT circuit using qiskit}  
\end{figure}
Note that there exist a built-in Qiskit QFT function that preforms a similar procedure as the defined QFT function above. However, it stars with building the circuit from qubits \textbf{upside down} and then swaps them afterwards. 
\subsection{Making measurements using the QFT}
\begin{figure}[H]
\centering{
\includegraphics[scale=0.31]{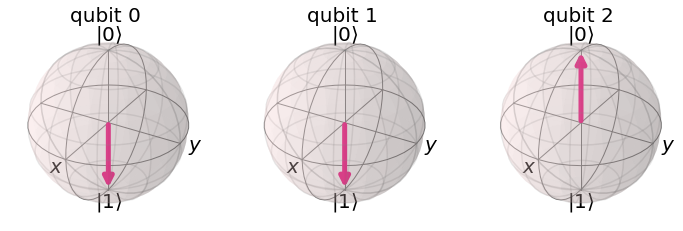}\\
\includegraphics[scale=0.31]{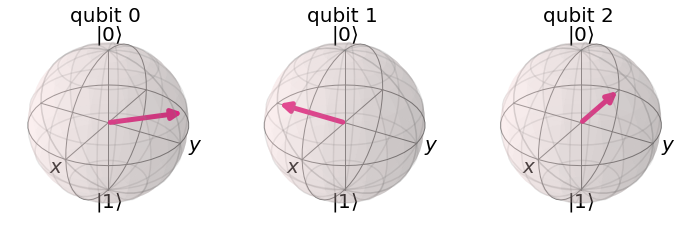}}
\caption{Bloch sphere representation of the QFT applied on the state $|6\rangle$}
\label{bloch2}
\end{figure}

After building the general QFT circuit, we can start using it for making different measurements.
We first consider applying the QFT to the state $|6\rangle$, and show the transformation from it's computational basis to its Fourier basis. The input in binary form is given by:
$$
|6\rangle= |110\rangle
$$
using Qiskit, \ref{bloch2} shows the Bloch sphere transformation of the state $|6\rangle$ from its computational basis to it's Fourier basis $|\tilde{6}\rangle$. 
\newline

\begin{figure}[!tbp]
  \centering
  \begin{subfigure}[b]{0.32\textwidth}
    \includegraphics[width=\textwidth]{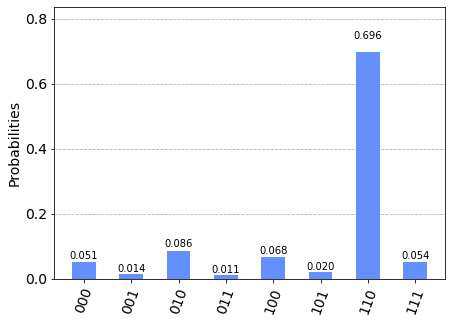}
    \caption{$|\tilde{6}\rangle$}
	\label{his1:a}
  \end{subfigure}
  \hfill
  \begin{subfigure}[b]{0.32\textwidth}
    \includegraphics[width=\textwidth]{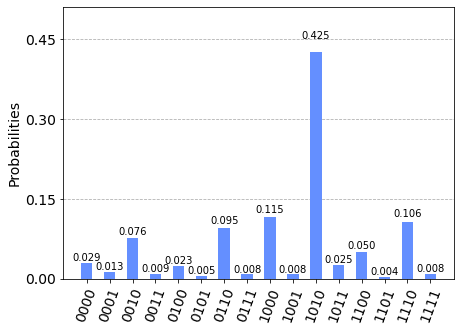}
    \caption{$|\tilde{10}\rangle$}
    \label{his1:b}
  \end{subfigure}
  \hfill
  \begin{subfigure}[b]{0.32\textwidth}
  \includegraphics[width=\textwidth]{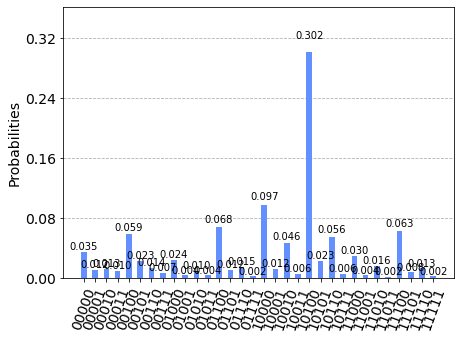}
  \caption{$|\tilde{20}\rangle$}
	\label{his1:c}
  \end{subfigure}
  \caption{Histogram of the inverse QFT on state (a)$|\tilde{6}\rangle$, (b)$|\tilde{10\rangle}$ and $|\tilde{20\rangle}$ }
\label{His1}	
\end{figure}

We can verify the prepared QFT function by applying an inverse QFT ($QFT^{\dagger}$) on the state $|\tilde{6}\rangle$ to obtain the state $|6\rangle$.
To prepare the state in the Fourier basis ($|\tilde{6}\rangle$),  we use the definition in equation \ref{eq:1}. 
\begin{figure}[H]
\raisebox{-0.2em}{\textcolor{cavitycolor}{\rule{1em}{1em}}} Python:
\begin{lstlisting}[caption={},language=Python] 
from qiskit import QuantumCircuit
import numpy as np
pi=np.pi
qc=QuantumCircuit(3)
qc.h(0)
qc.p(6*pi/4,0)
qc.h(1)
qc.p(6*pi/2,1)
qc.h(2)
qc.p(6*pi,2)
qc.draw()
\end{lstlisting}
\raisebox{-0.2em}{\textcolor{cavitycolor}{\rule{1em}{1em}}} Circuit: 
\begin{tikzpicture}
\begin{yquant}
qubit j[3];
h j[0];
box {$P_1$} j[0] ;
h j[1];
box {$P_2$} j[1] ;
h j[2];
box {$P_3$} j[2] ;
\end{yquant}
\end{tikzpicture}
\caption{Generation of an initial state $|6\rangle$ in the Fourier basis}
\end{figure}
Following the procedure laid out by Qiskit documentation, we use Qiskit's inverse QFT code, apply it to our circuit and load it in a real quantum device.  \ref{his1:a} Shows the histogram of the function $QFT^{\dagger}|\tilde{6}\rangle$ runs on a real quantum device which shows the state (110) as the highest probability outcome thus verifying the QFT function.

\begin{figure}[h!]
\centering{
\includegraphics[scale=0.31]{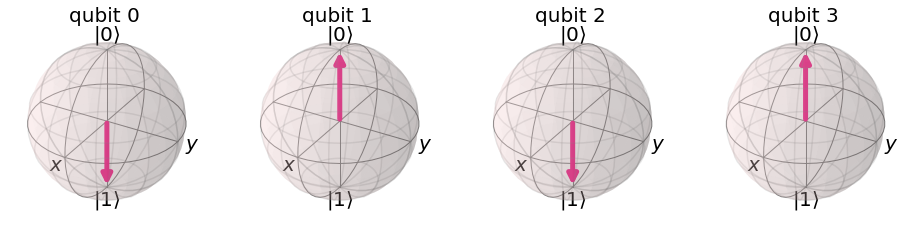}\\

\includegraphics[scale=0.31]{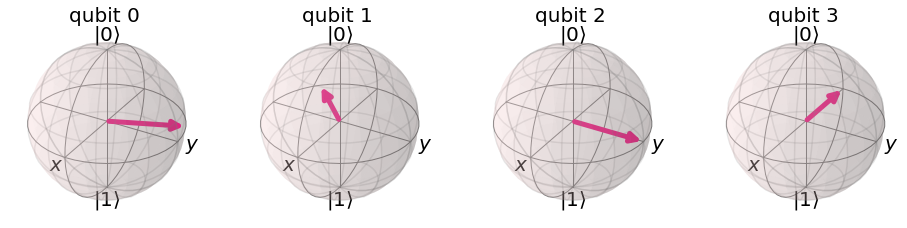}}
\caption{Bloch sphere representation of the QFT applied on the state $|10\rangle$}
\label{bloch3}
\end{figure}

We next consider the state $|10\rangle$:
$$
|10\rangle =|1010\rangle
$$
We follow similar procedure as before to find it's basis transformation in \ref{bloch3} and the histogram of $QFT^{\dagger}|\tilde{10}\rangle$. Where \ref{his1:b} shows the state (1010) to be with highest probability outcome. \ref{his1:c} also shows the outcome of $QFT^{\dagger}|\tilde{20}\rangle$ to be the state (10100)$=|20\rangle$. We observe that with increasing number of qubits in our quantum circuit, from 3 for the state $|6\rangle$ to 4 for the state $|10\rangle$ to 5 for the state $|20\rangle$, the probabilities get increasingly smaller. i.e \textbf{accuracy decreases with increasing number of qubits.}

\subsection{QFT and it's relationship to the FFT}
We have illustrated how we can use the QFT in a real quantum device. In this subsection we will look into the Fast Fourier Transform (FFT) and how it is closely relates to the QFT.
The Fast Fourier transform (FFT) is used for the fast computation of the DFT, which makes use of the symmetries in the DFT and can reduce the computational cost from O($N^2$) to O[$N log N]$ through factorizing the DFT matrix into two sparse matrices. 
From section \ref{sec:dft} we have derived a form for the DFT matrix for N=4 and is given by equation \ref{eq022}. 
\cite{Young} showed how the DFT matrix can be decomposed into two sparse matrices in the following way:
\begin{equation}
F_4=U_1U_2
\end{equation}
$$
\begin{bmatrix}
1 & 1 & 1 & 1\\
1 & i & -1 &  -i\\
1 & -1 & 1 & -1\\
1 & -i & -1 & i\\
\end{bmatrix}
=\begin{bmatrix}
1 & 0 & 1 & 0\\
0 &1 &0 &1\\
1 & 0 & i^2 & 0\\
0& 1 & 0 &i^2
\end{bmatrix}
 \begin{bmatrix}
1 & 1 & 0 & 0 \\
0 & 0 & 1 & i\\
1 & i^2& 0 & 0 \\
0 & 0 & 1 & i^3\\
\end{bmatrix}
$$

where $U_1$ and $U_2$ are the two sparse matrices used in the FFT computations. Moreover, we know that the QFT for a two qubit system can be constructed by applying a Hadamard gate on the first qubit ($H_1$) followed by a controlled phase $(R)$ gate, then another Hadamard gate on the second qubit $(H_2)$ and finally a swap gate $(S)$. In matrix representation, this is given by: $QFT=SH_2RH_1$

%
%
%
%

\cite{Young} showed that we can decompose the QFT into two sparse matrices in the following way:
\begin{equation}
U_2=H_1
\end{equation}

\begin{equation}
U_1=SH_2R
\end{equation}
\begin{equation}
QFT=U_1U_2
\end{equation}
Thus we can see that the FFT and the QFT are closely related.
\newline
We now consider applying the FFT on the signal $(0,1)$:
\begin{figure}[H]
\raisebox{-0.2em}{\textcolor{cavitycolor}{\rule{1em}{1em}}} Python:
\begin{lstlisting}[caption={},language=Python] 
#FFT computation of a signal
from scipy.fftpack import fft  
import numpy as np  
x = np.array([0,1,0,0])  
y = fft(x)  
print(y) 
\end{lstlisting}
\caption{A FFT computation of the signal (0,1,0,0) }
\end{figure}
This produces the results: [ 1.-0.j  0.-1.j -1.-0.j  0.+1.j].\newline

Next, using the Qiskit's Built-in QFT function, we find the QFT of the state $|1\rangle= |01\rangle$
\begin{figure}[H]
\raisebox{-0.2em}{\textcolor{cavitycolor}{\rule{1em}{1em}}} Python:
\begin{lstlisting}[caption={},language=Python] 
from qiskit import QuantumCircuit
from qiskit.circuit.library import QFT
qc = QuantumCircuit(2)  
qc.x(0)
#apply QFT
qc.append(QFT(2),[0,1])
results = qi.Statevector.from_instruction(qc)
sim = QasmSimulator()
options = {'method': 'statevector'}
execute(qc, sim, backend_options=options)
results.draw('latex', prefix='Statevector1:')
\end{lstlisting}
\caption{A generation of Qiskit's QFT on the state $|01\rangle$}
\end{figure}
which gives the output state:
$$
\frac{1}{2} |00\rangle-\frac{i}{2}|01\rangle -\frac{1}{2}|10\rangle -\frac{i}{2}|11\rangle =\begin{bmatrix}
1 \\
i \\
-1 \\
-i
\end{bmatrix}
$$
Which is the same result we got using the FFT.


\pagebreak
\section{USING THE QFT FOR SIGNAL PROCESSING}\label{sec:qft_sig_proc}
The QFT is analogous to the DFT by transforming information between the time and frequency domain. This means that we can use the QFT to find the frequencies of some sample input encoded on the basis states of the system. This is the basis of the study of signal processing.

\subsection{Using the QFT to Find a Musical Note}
\subsubsection{Generating a Musical Note}
To begin this experiment, first we will generate a single musical note. We have selected the note A4 otherwise known as ``A440'' as it is a pitch standard of a sound oscillating at exactly 440Hz. 

\begin{figure}[H]
\centering
\includegraphics[scale=0.15]{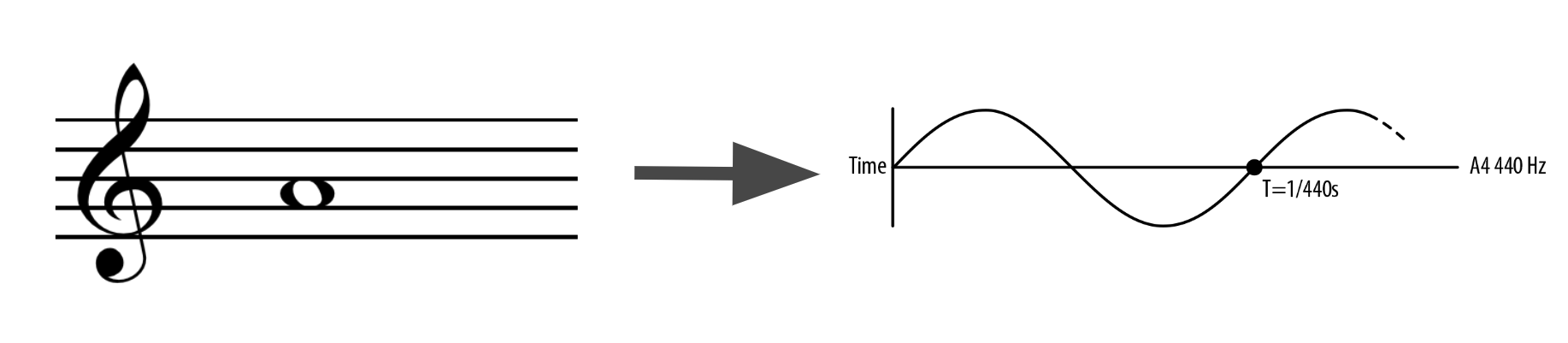}
\caption{The Musical Note A440}
\label{fig:a440}
\end{figure}

\subsubsection{Sampling Audio}
Audio is any sound or noise that the human ear is capable of hearing measured in units of cycles per second or Hertz (Hz). Since audio by nature is an analog signal, we must convert this analog signal to a digital signal by sampling at a rate fast enough to capture the wave. In signal processing, the Nyquist frequency is a characteristic of a sampler, which converts a continuous function or signal into a discrete sequence. The Nyquist rate is equal to twice the highest frequency (bandwidth) of a given function or signal.\cite{grenander1959probability}. Since our signal is 440Hz, this means we must sample at a rate of at least 880Hz to fully capture the signal without aliasing.
\newline
\newline
Fortunately, The WAV audio format developed by Microsoft stores audio at about 10 mega-bytes (MB) per minute at a 44.1 kHz sample rate using stereo 16-bit samples\cite{wav} which is far above the Nyquist frequency needed to capture this note. Once the note is recorded and the WAV file is created, we can use python to create an array of amplitudes for each sample and plot the results.  

\begin{figure}[H]
\raisebox{-0.2em}{\textcolor{cavitycolor}{\rule{1em}{1em}}} Python:
\begin{lstlisting}[caption={},language=Python] 
import math
import numpy as np
from pydub import AudioSegment
import matplotlib.pyplot as plt
from qiskit import QuantumCircuit, QuantumRegister, assemble, Aer
from qiskit.visualization import plot_histogram
#Turn wav audio file into python list of sample amplitudes
n_qubits = 10 #number of qubits that will be used to encode samples in QFT
audio = AudioSegment.from_file('440.wav') #store contents of audio
audio.set_frame_rate(10000) #reduce the frame rate from 44.1KHz to 10KHz
frame_rate = audio.frame_rate
samples = audio.get_array_of_samples()[:2**n_qubits] #create list of length 2^n_qubits to store samples
#Plot Samples
plt.xlabel('Sample #')
plt.ylabel('128-bit Amplitude Value')
plt.plot(list(range(len(samples))), samples)
\end{lstlisting}
\caption{Creating a List in Python of Audio Amplitudes} 
\end{figure}

\begin{figure}[H]
\centering
\includegraphics[scale=0.55]{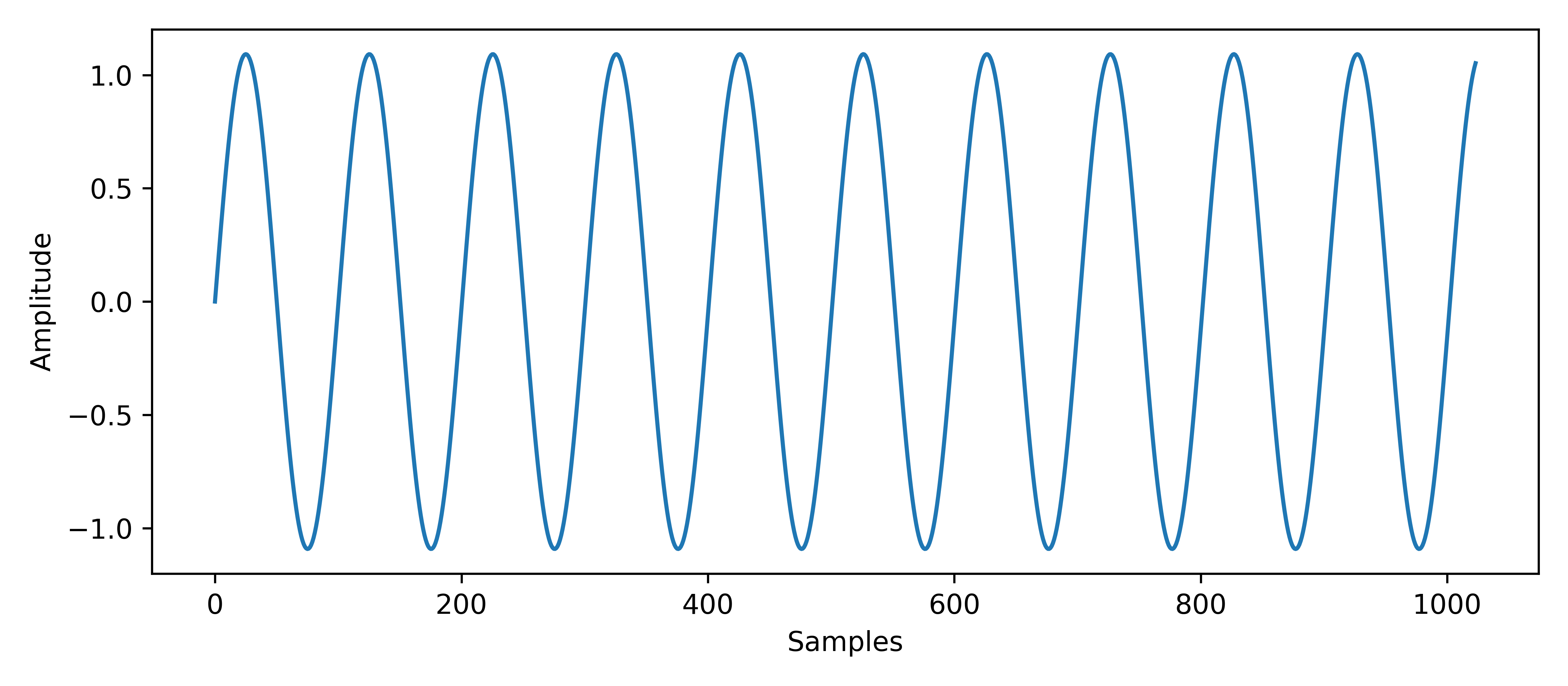}
\caption{Output Audio Waveform of A440}
\label{fig:samp_440}
\end{figure}

\subsubsection{Encoding Samples onto Qubits in the QFT}
To use QFT on the sampled wave, we must encode the amplitude of each sample into the amplitudes of the basis states of our quantum system. Since our QFT circuit is utilizing 10 qubits, we can encode up to $2^{10} =  1024$ amplitudes into each basis state of the system. 

\begin{equation}
    \large |\psi_{audio}\rangle = a_{0}|0000000000\rangle + a_{1}|0000000010\rangle + ...+ a_{1024}|1111111111\rangle
\end{equation}

Since $|\psi_{audio}\rangle$ is a quantum state, the sum of the amplitudes must be equal to 1. 
\begin{equation}
    \large \langle\psi_{audio}|\psi_{audio}\rangle = \sum_{i=0}^{1024}|a_{i}|^{2} = 1
\end{equation}
This means that we must normalize each amplitude so that this property is maintained and then encode each amplitude into each basis state of the system. Qiskit's 'initialize' function makes this process easy by using the Schmidt Decomposition and CNOT and single-qubit rotations to initialize the basis states we need. 

\begin{figure}[H]
\raisebox{-0.2em}{\textcolor{cavitycolor}{\rule{1em}{1em}}} Python:
\begin{lstlisting}[caption={},language=Python] 
#Encodes a normalized set of amplitudes from the audio samples onto the states of the qubits
def create_encoded_qc(samples):
    amplitudes = len(samples)
    num_qubits = int(math.log2(amplitudes))
    
    q = QuantumRegister(num_qubits)
    qc = QuantumCircuit(q)
    
    normalized_amplitudes = samples / np.linalg.norm(samples)
    qc.initialize(normalized_amplitudes, [q[i] for i in range(num_qubits)])
    return qc
#Create Quantum Circuit
audio_qc = create_encoded_qc(samples)
\end{lstlisting}
\caption{Encoding Normalized Amplitudes onto a Qiskit Circuit} 
\end{figure}

\subsubsection{Applying the QFT and Measuring}
The next step is to apply the QFT on our 10 qubits, and create a classical register at the end to store our measurement output. For now, we will simulate this using the Aer simulator provided by Qiskit. 

\begin{figure}[H]
\raisebox{-0.2em}{\textcolor{cavitycolor}{\rule{1em}{1em}}} Python:
\begin{lstlisting}[caption={},language=Python] 
#Apply QFT to quantum circuit, create measurement register, and draw visualization of circuit
qft(audio_qc, n_qubits)
audio_qc.measure_all()

#Simulation
sim = Aer.get_backend("aer_simulator")
qobj = assemble(audio_qc)
counts = sim.run(qobj).result().get_counts()
qft_counts = get_qft_counts(counts, n_qubits)[:len(samples)//2]
plot_histogram(counts)
\end{lstlisting}
\caption{Applying the QFT} 
\end{figure}

Since each qubit is measured in the Pauli-Z basis, all qubits will collapse into one of the two eigenstates of the Pauli-Z operator probabilities dictated by the superposition that we put the qubits in. The QFT on each qubit in a superposition has the effect of causing the qubits collapse in a way that gives us the Fourier transform of the input amplitudes.

\begin{figure}[H]
\centering
\includegraphics[scale=0.35]{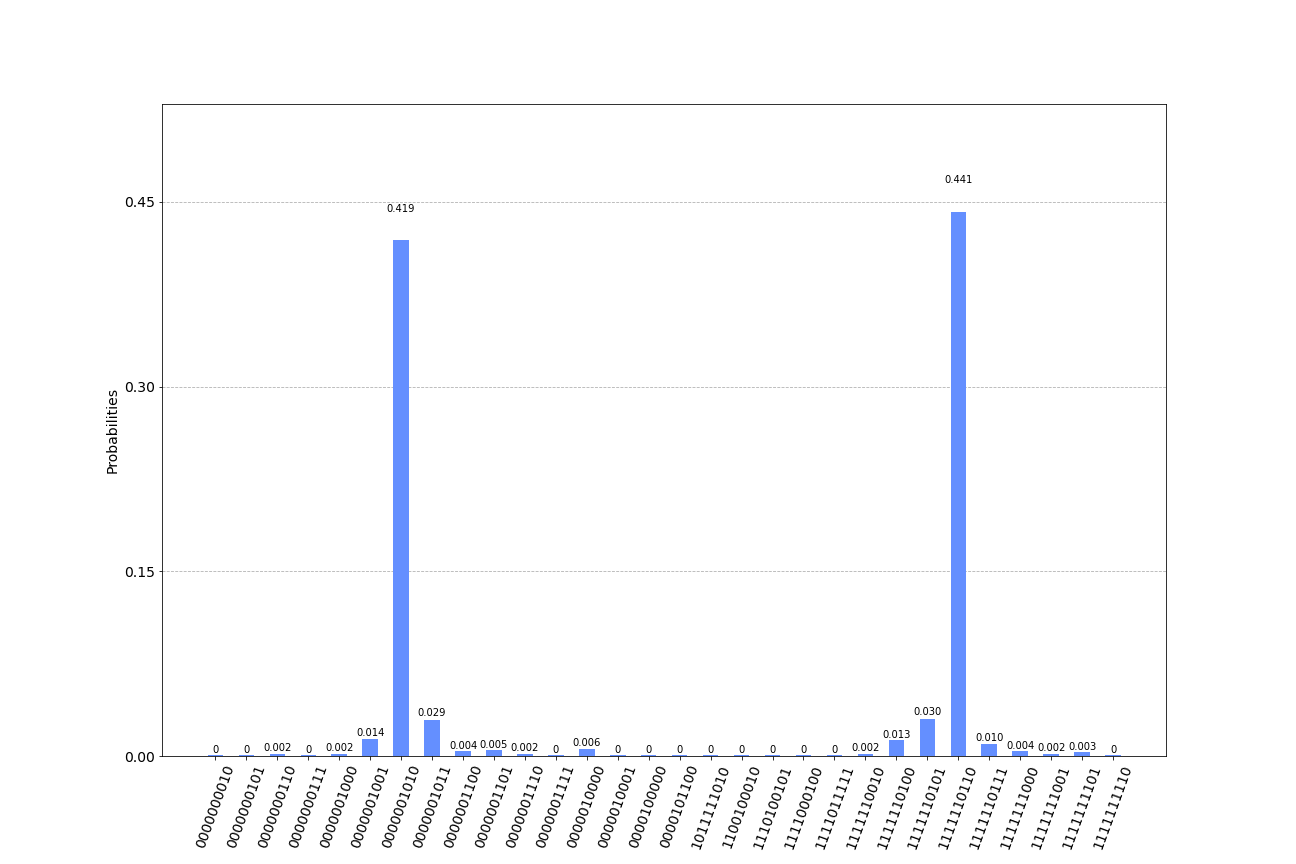}
\caption{Histogram of the QFT Output of A440}
\label{fig:qft_440}
\end{figure}

\subsubsection{Decoding the Result}
If the final output of the state is given by :
\begin{equation}\large |\psi\rangle = \sum_{i}a_{i}|i\rangle\end{equation}

where $i$ are the basis states, then the information we need to determine the frequency of the musical note is stored in the amplitude in $a_{i}$. Here we will use Python to evaluate the top indices of our output. To get the value of frequency we have to multiply these peaks by $\frac{frame\_rate}{2^n qubits}Hz$.

\begin{figure}[H]
\raisebox{-0.2em}{\textcolor{cavitycolor}{\rule{1em}{1em}}} Python:
\begin{lstlisting}[caption={},language=Python] 
#Calculate Frequencies and print top 2 frequencies
top_indices = np.argsort(-np.array(qft_counts))
freqs = top_indices*frame_rate/2**n_qubits
print(freqs[:2])
\end{lstlisting}
\caption{Obtaining the Frequency} 
\end{figure}

The top two indices computed for frequency were computed to be:
\begin{equation}
[430.6640625, \;\;   473.73046875]
\end{equation}
These were the top two frequencies output by our QFT circuit. These are very close to the actual frequency which we know to be 440Hz since we sampled the musical note A440. 

\subsection{Using the QFT on a Musical Chord}
Now that we have demonstrated using the QFT to find a single musical note, the next step is to determine if we can use this method to find multiple musical notes played together. Multiple notes played at the same time is known as a musical chord. Below is a F-major chord comprised of three notes: C3, F3, and A4 all played simultaneously. The resulting audio signal will be three frequencies in superposition. C3 corresponds to 130.81Hz, F3 to 174.61Hz, and A4 to 440.0Hz. 

\begin{figure}[H]
\centering
\includegraphics[scale=0.10]{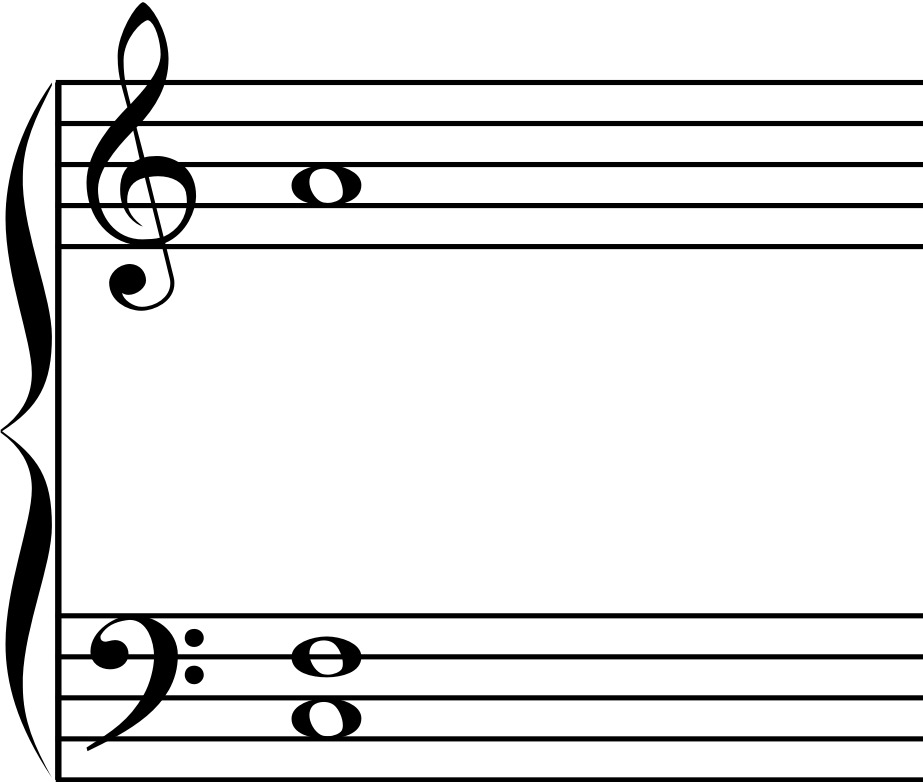}
\caption{F-Major Chord}
\label{fig:fmaj}
\end{figure}

Following the same steps on the single note example, we can sample the audio of this chord played and obtain an array of amplitudes for each sample and plot the results.

\begin{figure}[H]
\centering
\includegraphics[scale=0.6]{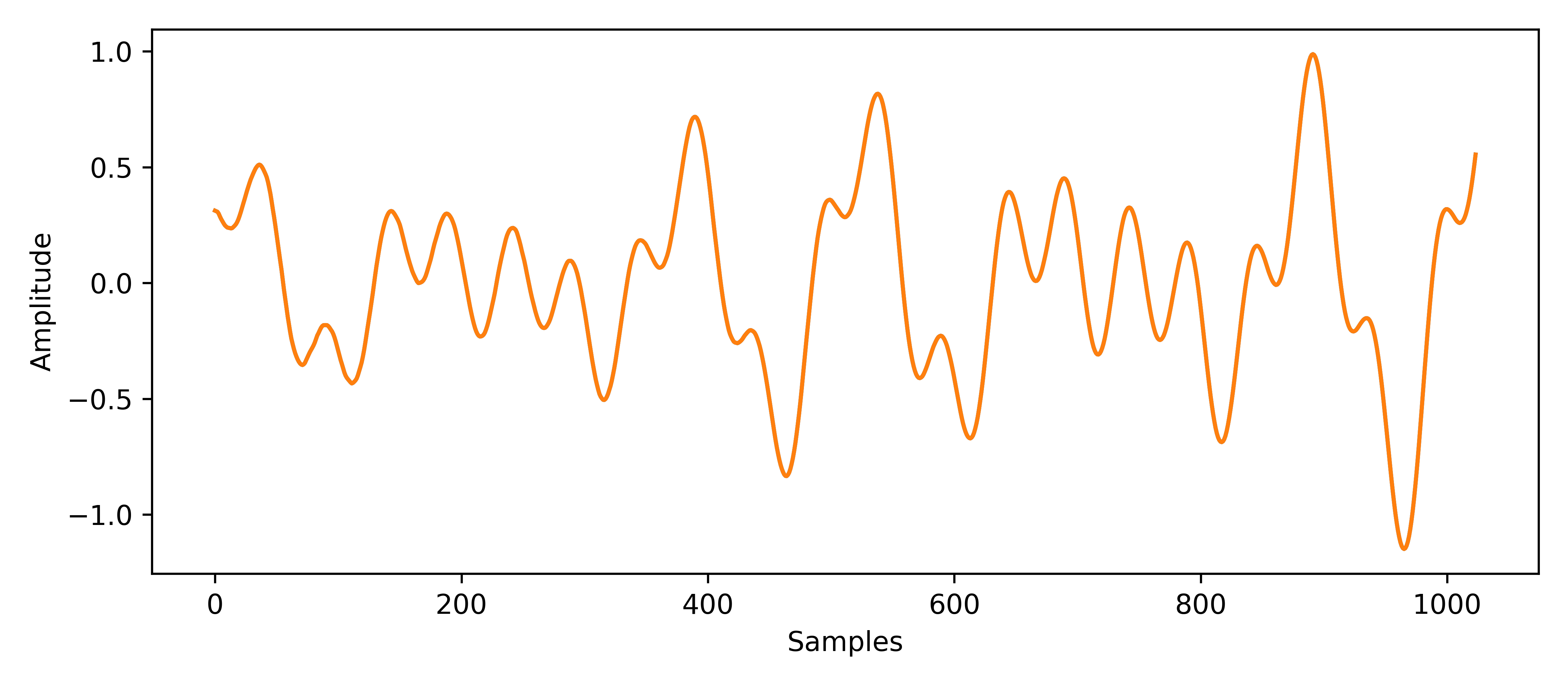}
\caption{Output Audio Waveform of F-Major Chord}
\label{fig:samp_fmaj}
\end{figure}

One might notice that it would be difficult to discern the three distinct frequencies of the musical chord by looking at the graph alone to calculate the period or frequency. The QFT is useful here because it will transform this data from the time domain to the frequency domain where we will see amplitudes of the three individual frequencies clearly. 
\newline
Next, we encode the samples again onto our quantum system and apply the QFT. After simulating the circuit and outcome we are left with the following results. 

\begin{figure}[H]
\centering
\includegraphics[scale=0.30]{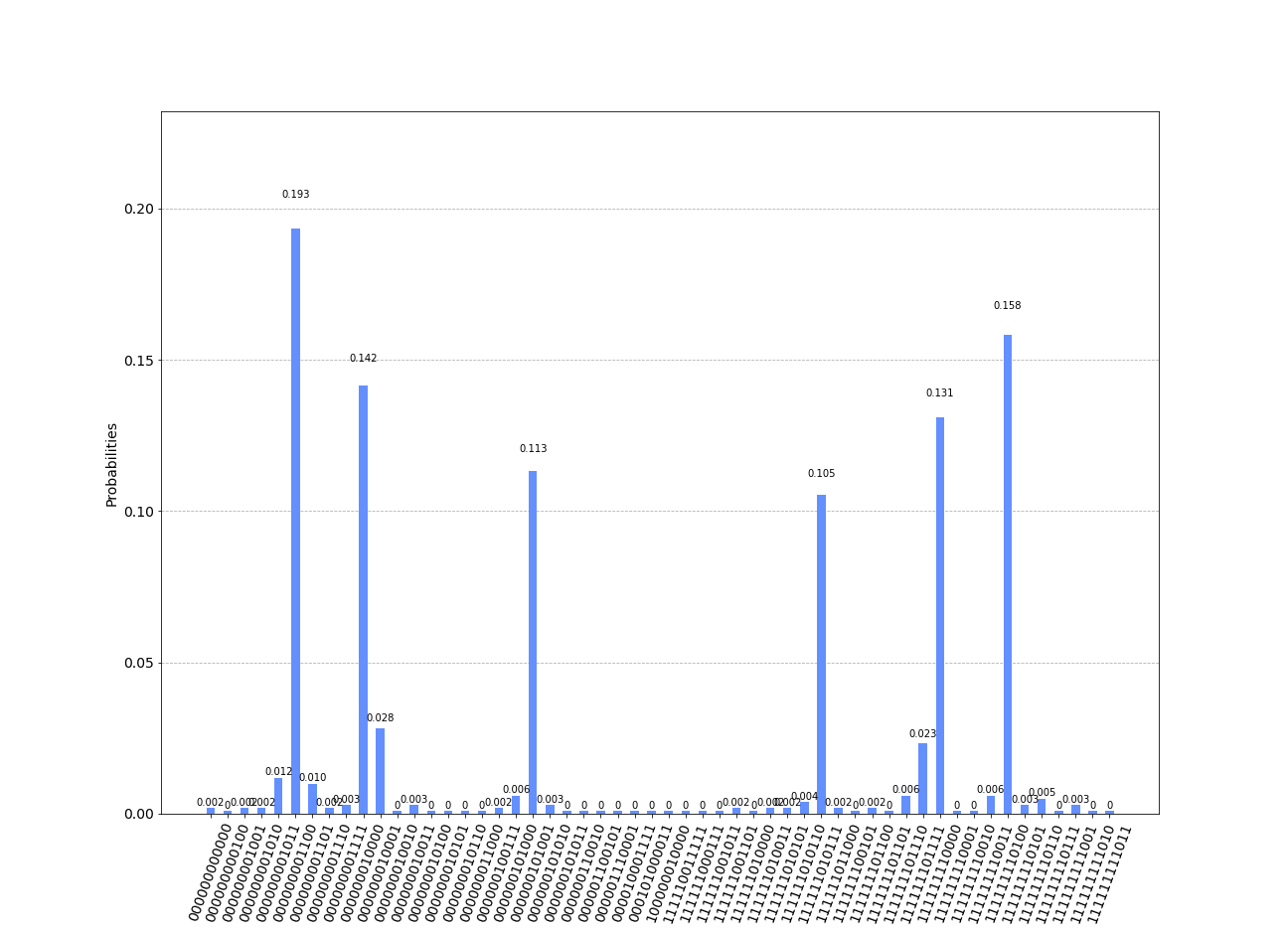}
\caption{Histogram of the F-Major QFT Output}
\label{fig:qft_fmaj}
\end{figure}

The top three indices of the histogram computed for frequency are:
\begin{equation}
[129.19921875,  \;\;    172.265625,   \;\;   441.43066406]
\end{equation}
This time, we will require the the top three frequencies output by our QFT circuit since we would like to know the frequencies of the three distinct notes in the F-major chord. The results are again, very close to our actual frequencies which we know to be 130.81Hz (C3), 174.61Hz (F3), 440.0Hz (A4).

\subsection{Using the QFT for Dual-Tone Multi-Frequency Decoding}
Now that we have demonstrated we can use the QFT to identify individual notes in a chord, we can demonstrate a real-world application of this technology. 
\newline
\newline
For this next experiment, we will be testing dual-tone multi-frequency signalling (DTMF) tones otherwise known as telephone ``dial-tones.'' These tones each correspond to a number on a keypad of a telephone and transmit a two frequencies simultaneously. One might recall an automated phone system where pressing a certain number accesses various menus or directs your call to the correct department\cite{itu}. Telephone systems that have this feature record the tone transmitted over the line when you press a specific dial-tone and use the real-time DFT to decode the audio into its frequency components which corresponds to the number pressed. It is a simple example of real-time signal processing technology used today.
\begin{center}
\begin{tabular}{ |p{2cm}||p{2cm}|p{2cm}|p{2cm}|  }
 \hline
 \multicolumn{4}{|c|}{\textbf{DTMF Keypad Frequencies}} \\ \hline
  & \textbf{1209 Hz} & \textbf{1336 Hz} & \textbf{1477 Hz}\\ \hline
 \textbf{697 Hz} & \centering 1 & \centering 2 &  3\\ \hline
 \textbf{770 Hz} & \centering 4 & \centering 5 &  6\\ \hline
 \textbf{852 Hz} & \centering 7 & \centering 8 &  9\\ \hline
 \textbf{941 Hz} & \centering * & \centering 0 & \#\\ \hline
\end{tabular}
\end{center}

Following the same steps as the musical chord example we can input a WAV file of DTMF corresponding to the number `1'. Below is an example of a combination of 1209Hz and 697Hz representing the number `1' on the keypad. 

\begin{figure}[H]
\centering
\includegraphics[scale=0.55]{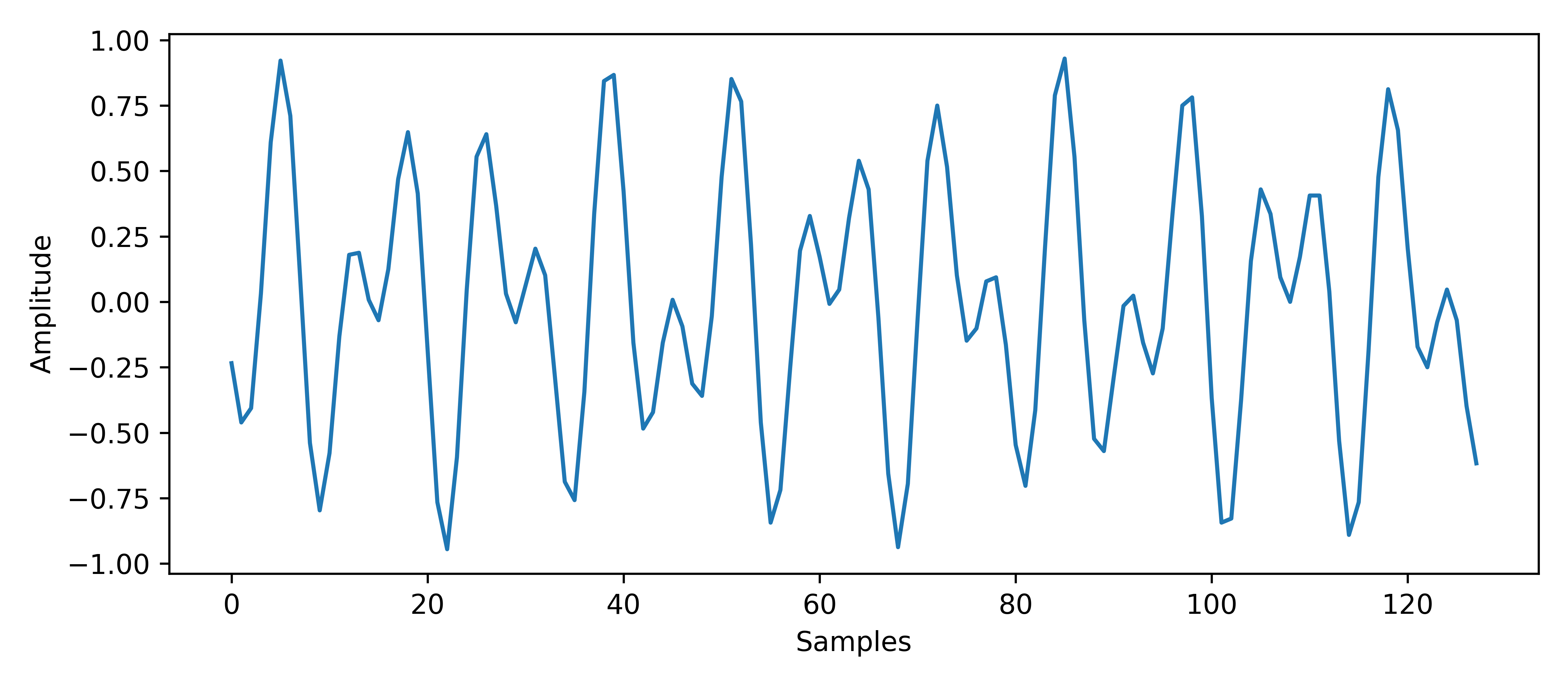}
\caption{Audio Waveform of 1209Hz and 697Hz (DTMF \#1)}
\label{fig:dtmf1}
\end{figure}

Much like the musical chord, the individual component frequencies of tone is difficult to discern by simply looking at the graph alone. The QFT will be used to transform this data from the time domain to the frequency domain where we will see amplitudes of the component frequencies clearly. We will apply the QFT, obtain the output histogram, and decode the highest indices again to find the component frequencies of the signal. 

\begin{figure}[H]
\centering
\includegraphics[scale=0.30]{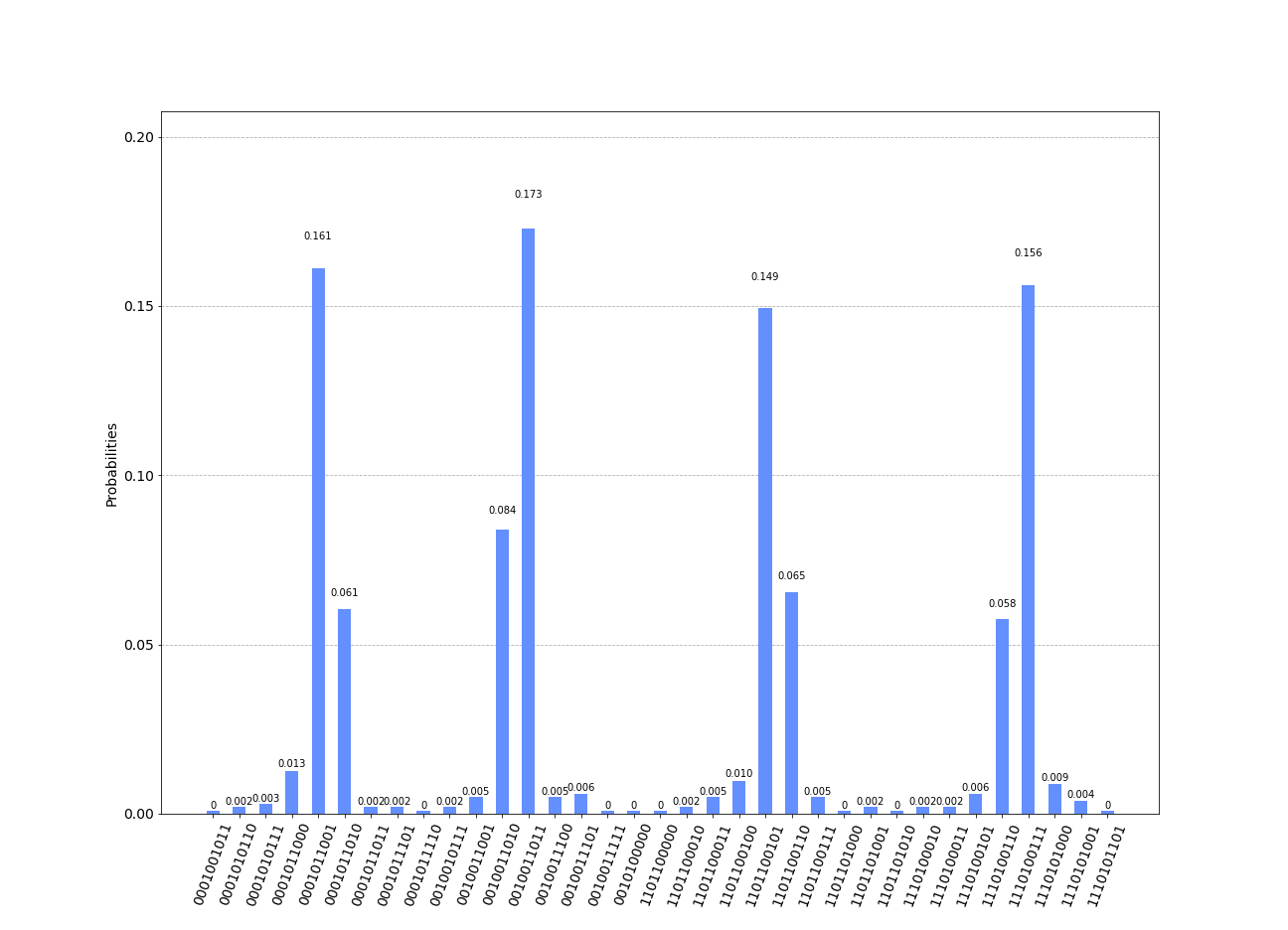}
\caption{Histogram of the DTMF '1' QFT Output}
\label{fig:qft_dtmf}
\end{figure}

The top frequencies computed were:
\begin{equation}
[1210.9375, \;\; 695.3125]
\end{equation}
The results are very close to our actual frequencies which we know to be 1209Hz and 697Hz which corresponds to the DTMF Keypad press for '1'. This process could be repeated if there were instead a string of multiple key-presses to decode a series of numbers. 

\subsection{Running the Experiment on Real Quantum Hardware}
So far, we have demonstrated the application for musical-note/frequency finding using a simulated quantum computer. Qiskit also has the ability to run quantum circuits on real quantum computers. IBM offers its quantum computers to run circuits in the cloud and return the results directly in python. The quantum computers available to us are limited to a low number of qubits, so we must modify our code to accommodate. It should be noted that using less qubits will directly reduce the number of samples we can take of our audio.
\newline
\newline
Nonetheless, we can try a 3 qubit example to find the note A440 once again using real quantum hardware. We will start by sampling $2^3 =  8$ times since we can only encode $2^3$ amplitudes into each basis state of a 3 qubit system. 

\begin{figure}[H]
\centering
\includegraphics[scale=0.6]{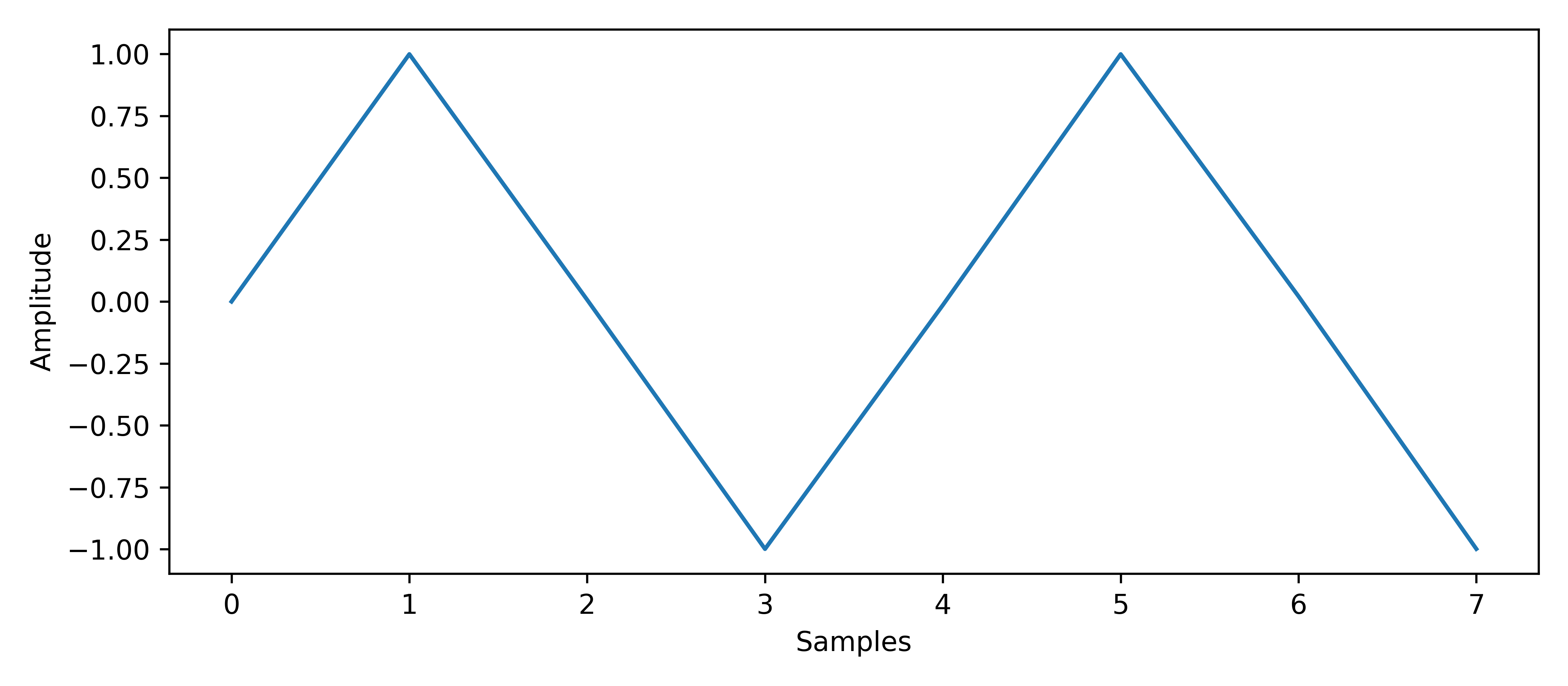}
\caption{A440 Sampled For 3 Qubits}
\label{fig:ibm_440}
\end{figure}

With fewer samples, one can see that the fidelity of the sine wave starts to deteriorate and begins to look more like a triangular wave. Next we can apply the QFT and run the circuit on one of IBM's real quantum computer and obtain a histogram to compute the frequency. 
\newline
\begin{figure}[H]
\raisebox{-0.2em}{\textcolor{cavitycolor}{\rule{1em}{1em}}} Python:
\begin{lstlisting}[caption={},language=Python] 
#IMBQ Setup
from qiskit import IBMQ, execute
IBMQ.enable_account('token')
provider = IBMQ.get_provider(hub='ibm-q')
from qiskit.providers.ibmq import least_busy
small_devices = provider.backends(filters=lambda x: x.configuration().n_qubits ==5
                                   and not x.configuration().simulator)
least_busy(small_devices)
#Program on Real Quantum Hardware
backend = provider.get_backend('ibmq_belem')
q_counts = execute(audio_qc, backend, shots=8192).result().get_counts()
ibmq_counts = get_qft_counts(q_counts, n_qubits)[:len(samples)//2]
plot_histogram(q_counts,figsize=(18, 13))
qtop_indices = np.argsort(-np.array(ibmq_counts))
freqs = qtop_indices*frame_rate/2**n_qubits
print(freqs[:1])
\end{lstlisting}
\caption{Running the Circuit on IBMQ} 
\end{figure}

\begin{figure}[H]
\centering
\includegraphics[scale=0.30]{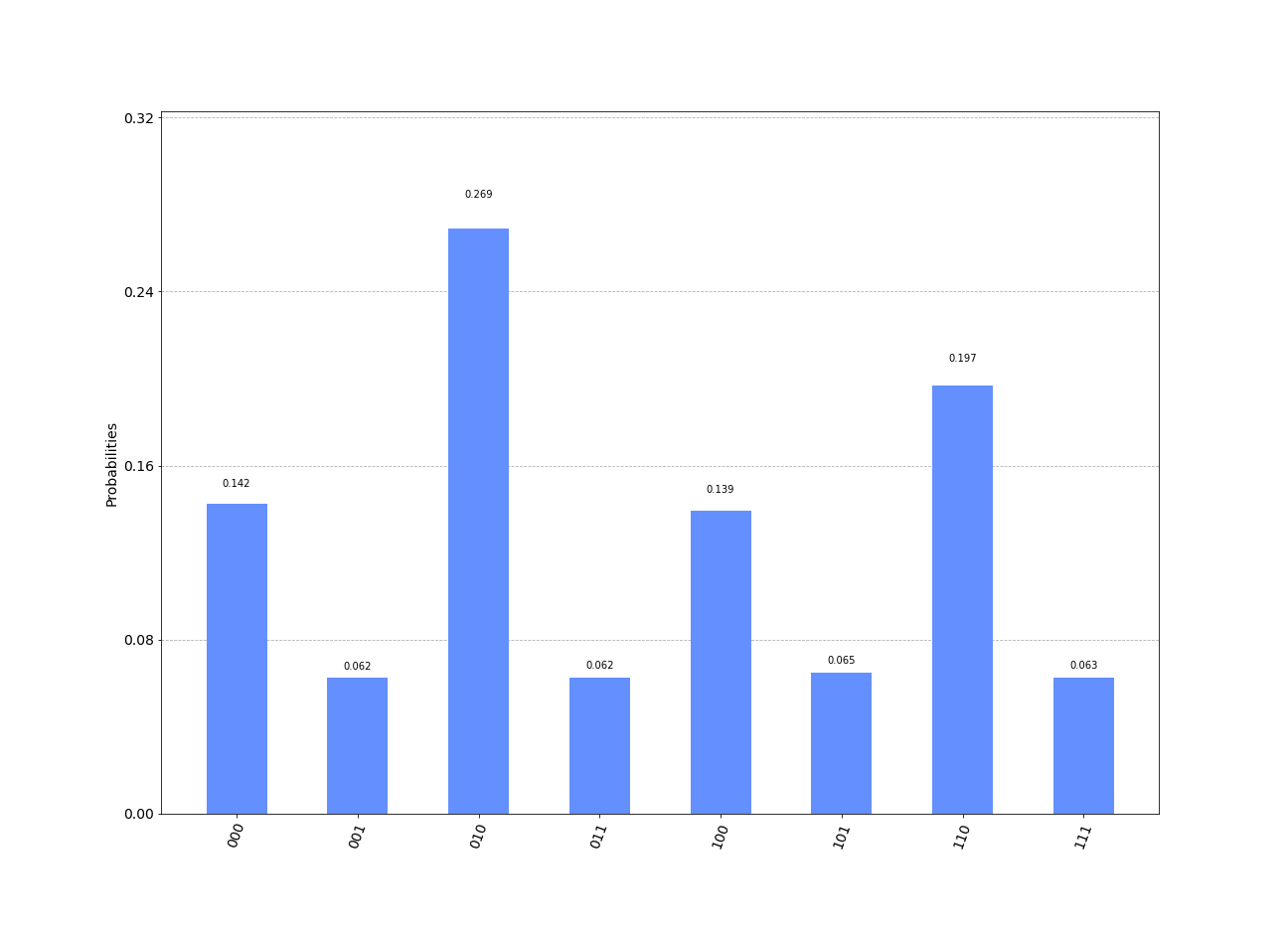}
\caption{Real Quantum Histogram of the QFT on A440}
\label{fig:qft_440}
\end{figure}

One thing to note is that there are more noisy states appearing in the output. One can see that several other encoded amplitudes are much higher when compared to the simulation histograms. Nonetheless, we can still take our highest index which happens to by the state $|010\rangle$. Computed for frequency using the same process as before, the frequency output is:
\begin{equation}
[441.0]
\end{equation}
\newline
The results are very close to the actual frequency known to be 440Hz.
\newline
\newline
The results of this particular experiment with three qubits is promising; however, a few things should be noted about the scalability of this application. Real quantum hardware tends to be noisy, especially when the number of gates becomes large. While the complexity of the QFT is known, it should also be noted that there must be a number of gates applied prior to the QFT to encode the amplitudes of our sample into basis states of the quantum system. The amount of gates for encoding scales with the number of n-qubits at $2^n$. While we were able to demonstrate this in a simulation without accounting for noise, a larger real quantum system would likely be prone to gate noise for encoding frequencies in addition to any noise that occurs with the QFT application. In simulation, more samples is equivalent to better accuracy of finding the frequency, but on real-quantum hardware, more qubits can ultimately introduce more noise due to added gates for encoding samples and the added gates for expanding the QFT. Perhaps with more research and future advances in quantum computing, we can begin to scale the QFT application for signal processing applications that are larger in scale and introduce error correction to this application.  

\subsection{Future Signal Processing Applications}
While this study offers a simple proof-of-concept for using the QFT for signal processing audio files, this proof can ultimately be extended to other signal processing applications. Signal processing by use of the QFT is feasible and could be used for improvements in computational complexity over classical computers. Ideally, quantum computers have the opportunity to make improvements in efficiency in the field of signal processing via the QFT. 


\section{CONCLUSIONS}\label{sec:outlook}

In this paper, we reviewed the quantum Fourier transform, which is crucial to many algorithms used in quantum processing. The QFT may be realized utilizing efficient tensor products of quantum operators as a logical extension of the discrete Fourier transform. We described that Quantum circuit verification can be approached in two distinct ways. The gates can be formally and logically verified using a purpose-built tool such as Cog \cite{hietala2019verified}, or they can be directly targeted using a symbolic mathematical description \cite{sympy2012}. The downside of the later approach is that it lacks Cog's higher-level abstractions, making it unsuitable for larger, more sophisticated quantum circuits. We have also used Qiskit to demonstrate the QFT by building a simple quantum circuit and using it to transform the basis from its computational basis to it's Fourier basis. Next, we built a general n-qubit QFT circuit and verified the implementation. Moreover, we used the inverse QFT to make measurements where we demonstrated that the accuracy of measurement decreases with increasing the number of qubits in our quantum circuit. Then we briefly looked into the relation between the QFT and the FFT. Finally, we formulated our note-detection algorithm using the QFT to offer a proof-of-concept to use the QFT for signal-processing. 
\newline
\newline
Future study will focus on expanding the use of quantum computing for signal processing applications beyond musical note detection and audio. The reduced computational complexity of the QFT versus the DFT is likely to provide a benefit in a future where quantum computing is more ubiquitous.


\section*{DISCLOSURE STATEMENT}
The authors are not aware of any affiliations, memberships, funding, or financial holdings that
might be perceived as affecting the objectivity of this review.

\section*{ACKNOWLEDGMENTS}
The authors gratefully acknowledge input and comments from Dr. Dave Clader, JHU.

\newpage
\listoffigures
\listoftables
\printindex 
\bibliography{refs}
\end{document}